\documentclass[a4paper, oneside, twocolumn, notitlepage, 10pt]{extarticle_ecoc}
\usepackage{ecoc}
\usepackage{amsmath,amssymb}
\usepackage{pgfplotstable}
\usepackage{subcaption}
\usepackage[capitalise]{cleveref}
\usepackage{siunitx}
\usepackage{url}
\usepackage{sfmath}

\definecolor{C1}{RGB}{031, 119, 180} 
\definecolor{C2}{RGB}{255, 127, 014} 
\definecolor{C3}{RGB}{044, 160, 044} 
\definecolor{C4}{RGB}{215, 039, 040} 
\definecolor{C6}{RGB}{148, 103, 189} 
\definecolor{C100}{RGB}{140, 086, 075} 
\definecolor{C7}{RGB}{227, 119, 194} 
\definecolor{C8}{RGB}{127, 127, 127} 
\definecolor{C9}{RGB}{188, 189, 034} 
\definecolor{C5}{RGB}{023, 190, 207} 

\definecolor{C11}{RGB}{174, 199, 232} 
\definecolor{C12}{RGB}{255, 187, 120} 
\definecolor{C13}{RGB}{152, 223, 138} 
\definecolor{C14}{RGB}{255, 152, 150} 
\definecolor{C15}{RGB}{197, 176, 213} 
\definecolor{C16}{RGB}{196, 156, 148} 
\definecolor{C17}{RGB}{247, 182, 210} 
\definecolor{C18}{RGB}{199, 199, 199} 
\definecolor{C19}{RGB}{219, 219, 141} 
\definecolor{C20}{RGB}{158, 218, 229} 
\usepackage[font=small]{caption}
\usetikzlibrary{backgrounds}
\usepackage{import}
\usepackage{titlesec}
\usepackage{tikz}
\usepackage{import}

\usetikzlibrary{positioning}
\usetikzlibrary{fit}
\usetikzlibrary{calc}
\usetikzlibrary{shapes}
\usetikzlibrary{math}
\usetikzlibrary{fpu}
\usetikzlibrary{decorations.markings}
\usetikzlibrary{decorations.pathreplacing}
\usetikzlibrary{arrows.meta}
\usetikzlibrary{intersections}

\makeatletter

\tikzset{outer sep=0}
\tikzset{inner sep=0}

\def\pgfaddtoshape#1#2{
	\begingroup
	\def\pgf@sm@shape@name{#1}%
	\let\anchor\pgf@sh@anchor
	#2%
	\endgroup
}

\newcommand{\anchorlet}[2]{
	\global\expandafter
	\let\csname pgf@anchor@\pgf@sm@shape@name @#1\expandafter\endcsname
	\csname pgf@anchor@\pgf@sm@shape@name @#2\endcsname
}

\pgfmathsetmacro{\NODESIZE}{42}
\pgfmathsetmacro{\NODETHICKNESS}{1.0}
\pgfmathsetmacro{\ROUNDEDCORNERS}{0.5mm}
\tikzset{node distance=0.5*\NODESIZE pt}

\def\FNODESIZE{\NODESIZE pt}

\def\QNODESIZE{0.25*\NODESIZE pt}

\tikzstyle{textstyle} = [text height=1.5ex, text depth=.5ex]
\tikzset{every label/.style=textstyle}

\tikzstyle{linestyle} = [line width = \NODETHICKNESS, rounded corners = \ROUNDEDCORNERS]
\tikzstyle{arrowstyle} = [>=stealth, linestyle]
\tikzstyle{<--} = [<-, arrowstyle]
\tikzstyle{-->} = [->, arrowstyle]
\tikzstyle{->-} = [linestyle, decoration={markings,	mark=at position 0.5 with {\arrow[arrowstyle]{>}}}, postaction={decorate}] 
\tikzstyle{-<-} = [linestyle, decoration={markings,	mark=at position 0.5 with {\arrow[arrowstyle]{<}}}, postaction={decorate}] 

\tikzset{   
    -A-/.style args={#1}{%
        linestyle, 
        decoration={markings, mark=at position #1 with {\arrow[>=Triangle, scale=.025*\NODESIZE]{>}}},
        postaction={decorate}
    },
    -A-/.default = {0.75}
}

\tikzset{   
    -AA-/.style args={#1, #2, #3}{%
        linestyle,
        decoration={markings, mark=at position #1 with {
            \node[amp={color=#2, width=#3, height=#3}, rotate=\pgfdecoratedangle] at (0,0) (inline_amp) {};
        }},
        postaction={decorate}
    },
    -AA-/.default = {0.5, E, 0.15}
}

\tikzstyle{<-->} = [<->, arrowstyle]
\tikzstyle{---} = [arrowstyle]

\tikzset{
    -PC-/.style args={#1}{
        linestyle, 
        decoration={markings, mark=at position #1 with {
            \draw[---,FO]
                (.05*\NODESIZE*\pgflinewidth,.05*\NODESIZE*\pgflinewidth) circle[radius=.05*\NODESIZE*\pgflinewidth];
            \draw[---,FO]
                (-.05*\NODESIZE*\pgflinewidth,.05*\NODESIZE*\pgflinewidth) circle[radius=.05*\NODESIZE*\pgflinewidth];
            \draw[---,FO]
                (0,-.05*\NODESIZE*\pgflinewidth) circle[radius=.05*\NODESIZE*\pgflinewidth];}},
        postaction={decorate}
        },
        -PC-/.default = {0.5}
}

\definecolor{C0}{RGB}{031, 119, 180} 
\definecolor{C1}{RGB}{031, 119, 180} 
\definecolor{C2}{RGB}{255, 127, 014} 
\definecolor{C3}{RGB}{044, 160, 044} 
\definecolor{C4}{RGB}{215, 039, 040} 
\definecolor{C6}{RGB}{148, 103, 189} 
\definecolor{C100}{RGB}{140, 086, 075} 
\definecolor{C7}{RGB}{227, 119, 194} 
\definecolor{C8}{RGB}{127, 127, 127} 
\definecolor{C9}{RGB}{188, 189, 034} 
\definecolor{C5}{RGB}{023, 190, 207} 

\definecolor{C0l}{RGB}{174, 199, 232} 
\definecolor{C11}{RGB}{174, 199, 232} 
\definecolor{C12}{RGB}{255, 187, 120} 
\definecolor{C13}{RGB}{152, 223, 138} 
\definecolor{C14}{RGB}{255, 152, 150} 
\definecolor{C15}{RGB}{197, 176, 213} 
\definecolor{C16}{RGB}{196, 156, 148} 
\definecolor{C17}{RGB}{247, 182, 210} 
\definecolor{C18}{RGB}{199, 199, 199} 
\definecolor{C19}{RGB}{219, 219, 141} 
\definecolor{C20}{RGB}{158, 218, 229} 

\definecolor{Snow}{HTML}{FBFBFB} 			
\definecolor{TUeRed}{RGB}{200, 25, 25}		
\definecolor{TUeGreen}{RGB}{25, 200, 113}	
\definecolor{TUeBlue}{RGB}{25, 113, 200}	

\definecolor{O}{RGB}{031, 119, 180} 	
\definecolor{Ol}{RGB}{174, 199, 232} 	
\definecolor{E}{RGB}{255, 127, 014} 	
\definecolor{El}{RGB}{255, 187, 120} 	
\definecolor{D}{RGB}{148, 103, 189}     
\definecolor{Dl}{RGB}{197, 176, 213} 	
\definecolor{EO}{RGB}{215, 039, 040} 	
\definecolor{EOl}{RGB}{255, 152, 150}   

\tikzstyle{FW} = [fill=white]			
\tikzstyle{FB} = [fill=white]			
\tikzstyle{FO} = [fill=C0l, draw=C0]	
\tikzstyle{FE} = [fill=C1l, draw=C1]	
\tikzstyle{FD} = [fill=C2l, draw=C2]	

\def\direce{e}
\def\direcw{w}
\def\direcn{n}
\def\direcs{s}

\def\flipfalse{0}
\newcount\portcount
\pgfdeclareshape{basic}{
	\inheritsavedanchors[from=rectangle] 
	
	\inheritanchor[from=rectangle]{center}
	\inheritanchor[from=rectangle]{mid}		
	\inheritanchor[from=rectangle]{base}	
	\inheritanchor[from=rectangle]{north}
	\inheritanchor[from=rectangle]{south}		
	\inheritanchor[from=rectangle]{west}		
	\inheritanchor[from=rectangle]{mid west}				
	\inheritanchor[from=rectangle]{base west}		
	\inheritanchor[from=rectangle]{north west}		
	\inheritanchor[from=rectangle]{south west}		
	\inheritanchor[from=rectangle]{east}
	\inheritanchor[from=rectangle]{mid east}
	\inheritanchor[from=rectangle]{base east}	
	\inheritanchor[from=rectangle]{north east}
	\inheritanchor[from=rectangle]{south east}	
	
	\inheritanchorborder[from=rectangle]	

	\inheritbackgroundpath[from=rectangle]	
}

\pgfaddtoshape{basic}{
	\anchor{west south west}{
		\pgf@process{\northeast}
		\pgf@ya=.5\pgf@y
		\pgf@process{\southwest}
		\pgf@y=1.5\pgf@y
		\advance\pgf@y by \pgf@ya
		\pgf@y=.5\pgf@y
	}
	\anchor{west north west}{
		\pgf@process{\northeast}
		\pgf@ya=1.5\pgf@y
		\pgf@process{\southwest}
		\pgf@y=.5\pgf@y
		\advance\pgf@y by \pgf@ya
		\pgf@y=.5\pgf@y
	}
	\anchor{east north east}{
		\pgf@process{\southwest}
		\pgf@ya=.5\pgf@y
		\pgf@process{\northeast}
		\pgf@y=1.5\pgf@y
		\advance\pgf@y by \pgf@ya
		\pgf@y=.5\pgf@y
	}
	\anchor{east south east}{
		\pgf@process{\southwest}
		\pgf@ya=1.5\pgf@y
		\pgf@process{\northeast}
		\pgf@y=.5\pgf@y
		\advance\pgf@y by \pgf@ya
		\pgf@y=.5\pgf@y
	}
	\anchor{north north west}{
		\pgf@process{\southwest}
		\pgf@xa=1.5\pgf@x
		\pgf@process{\northeast}
		\pgf@x=.5\pgf@x
		\advance\pgf@x by \pgf@xa
		\pgf@x=.5\pgf@x
	}
	\anchor{north north east}{
		\pgf@process{\southwest}
		\pgf@xa=.5\pgf@x
		\pgf@process{\northeast}
		\pgf@x=1.5\pgf@x
		\advance\pgf@x by \pgf@xa
		\pgf@x=.5\pgf@x
	}
	\anchor{south south west}{
		\pgf@process{\northeast}
		\pgf@xa=.5\pgf@x
		\pgf@process{\southwest}
		\pgf@x=1.5\pgf@x
		\advance\pgf@x by \pgf@xa
		\pgf@x=.5\pgf@x
	}
	\anchor{south south east}{
		\pgf@process{\northeast}
		\pgf@xa=1.5\pgf@x
		\pgf@process{\southwest}
		\pgf@x=.5\pgf@x
		\advance\pgf@x by \pgf@xa
		\pgf@x=.5\pgf@x
	}
}

\pgfaddtoshape{basic}{
	\anchorlet{c}{center}		
	\anchorlet{n}{north}
	\anchorlet{e}{east}
	\anchorlet{s}{south}
	\anchorlet{w}{west}			
	\anchorlet{se}{south east}
	\anchorlet{sw}{south west}
	\anchorlet{ne}{north east}
	\anchorlet{nw}{north west}
	\anchorlet{wsw}{west south west}
	\anchorlet{wnw}{west north west}
	\anchorlet{ene}{east north east}
	\anchorlet{ese}{east south east}
	\anchorlet{nnw}{north north west}
	\anchorlet{nne}{north north east}
	\anchorlet{ssw}{south south west}
	\anchorlet{sse}{south south east}
}	

\pgfdeclareshape{ampshape}{
	\savedmacro\direction{
		\edef\direction{\pgfkeysvalueof{/tikz/ampkeys/direction}}%
	}
	\saveddimen\minwidth{
		\pgfmathsetlength\pgf@x{\pgfshapeminwidth}%
	}
	\saveddimen\minheight{
		\pgfmathsetlength\pgf@x{\pgfshapeminheight}%
	}
	\inheritsavedanchors[from=basic] 
	
	\inheritanchor[from=basic]{center}
	\inheritanchor[from=basic]{mid}		
	\inheritanchor[from=basic]{base}	
	\inheritanchor[from=basic]{north}
	\inheritanchor[from=basic]{south}		
	\inheritanchor[from=basic]{west}		
	\inheritanchor[from=basic]{mid west}				
	\inheritanchor[from=basic]{base west}		
	\inheritanchor[from=basic]{north west}		
	\inheritanchor[from=basic]{south west}		
	\inheritanchor[from=basic]{east}
	\inheritanchor[from=basic]{mid east}
	\inheritanchor[from=basic]{base east}	
	\inheritanchor[from=basic]{north east}
	\inheritanchor[from=basic]{south east}
	\inheritanchor[from=basic]{west south west}%
	\inheritanchor[from=basic]{west north west}%
	\inheritanchor[from=basic]{east north east}%
	\inheritanchor[from=basic]{east south east}%
	\inheritanchor[from=basic]{north north west}%
	\inheritanchor[from=basic]{north north east}%
	\inheritanchor[from=basic]{south south west}%
	\inheritanchor[from=basic]{south south east}%
	
	\inheritanchorborder[from=basic]	
	\inheritbackgroundpath[from=basic]

    \pgfutil@g@addto@macro\pgf@sh@s@ampshape{%
        \pgfutil@ifundefined{pgf@anchor@ampshape@in0}{
	        \expandafter\xdef\csname pgf@anchor@ampshape@in0\endcsname{%
	            \noexpand\ampshape@port{0}
	        }%
	    }{}%
        \pgfutil@ifundefined{pgf@anchor@ampshape@in}{
	        \expandafter\xdef\csname pgf@anchor@ampshape@in\endcsname{%
	            \noexpand\ampshape@port{0}
	        }%
	    }{}%
        \pgfutil@ifundefined{pgf@anchor@ampshape@out0}{
	        \expandafter\xdef\csname pgf@anchor@ampshape@out0\endcsname{%
	            \noexpand\ampshape@port{1}
	        }%
	    }{}%
        \pgfutil@ifundefined{pgf@anchor@ampshape@out}{
	        \expandafter\xdef\csname pgf@anchor@ampshape@out\endcsname{%
	            \noexpand\ampshape@port{1}
	        }%
	    }{}%
	}
}

\def\ampshape@port#1{
    \northeast	

    \ifnum#1=0	
	    \if\direction\direce
			\pgf@x=-\pgf@x
		    \pgf@ya= \pgf@y
		    \pgfmathsetlength{\pgf@y}{\pgf@ya-0.5*\minheight}%
		\fi
	    \if\direction\direcw
			\pgf@x=\pgf@x
		    \pgf@ya= \pgf@y
		    \pgfmathsetlength{\pgf@y}{\pgf@ya-0.5*\minheight}%
		\fi
	    \if\direction\direcn
			\pgf@y=-\pgf@y
		    \pgf@xa=\pgf@x
		    \pgfmathsetlength{\pgf@x}{\pgf@xa-0.5*\minwidth}%
		\fi
	    \if\direction\direcs
			\pgf@y=\pgf@y
		    \pgf@xa= \pgf@x
		    \pgfmathsetlength{\pgf@x}{\pgf@xa-0.5*\minwidth}%
		\fi
	\else	
	    \if\direction\direce
			\pgf@x=\pgf@x
		    \pgf@ya= \pgf@y
		    \pgfmathsetlength{\pgf@y}{\pgf@ya-0.5*\minheight}%
		\fi
	    \if\direction\direcw
			\pgf@x=-\pgf@x
		    \pgf@ya= \pgf@y
		    \pgfmathsetlength{\pgf@y}{\pgf@ya-0.5*\minheight}%
		\fi
	    \if\direction\direcn
			\pgf@y=\pgf@y
		    \pgf@xa= \pgf@x
		    \pgfmathsetlength{\pgf@x}{\pgf@xa-0.5*\minwidth}%
		\fi
	    \if\direction\direcs
			\pgf@y=-\pgf@y
		    \pgf@xa= \pgf@x
		    \pgfmathsetlength{\pgf@x}{\pgf@xa-0.5*\minwidth}%
		\fi
	\fi
}

\pgfaddtoshape{ampshape}{
	\anchorlet{c}{center}		
	\anchorlet{n}{north}
	\anchorlet{e}{east}
	\anchorlet{s}{south}
	\anchorlet{w}{west}			
	\anchorlet{se}{south east}
	\anchorlet{sw}{south west}
	\anchorlet{ne}{north east}
	\anchorlet{nw}{north west}
	\anchorlet{wsw}{west south west}
	\anchorlet{wnw}{west north west}
	\anchorlet{ene}{east north east}
	\anchorlet{ese}{east south east}
	\anchorlet{nnw}{north north west}
	\anchorlet{nne}{north north east}
	\anchorlet{ssw}{south south west}
	\anchorlet{sse}{south south east}
}

\tikzset{
	/tikz/ampkeys/.cd,
	height/.initial=0.5,
	width/.initial=0.5,
	color/.initial=O,
	direction/.initial=e,
	linestyle/.initial={linestyle, rounded corners = 0},
	/tikz/amp/.code={
		\pgfqkeys{/tikz/ampkeys}{#1}%
		\tikzset{/tikz/ampkeys/drawer/.expanded=%
			{\pgfkeysvalueof{/tikz/ampkeys/direction}}%
			{\pgfkeysvalueof{/tikz/ampkeys/height}}%
			{\pgfkeysvalueof{/tikz/ampkeys/width}}%
			{\pgfkeysvalueof{/tikz/ampkeys/color}}%
			{\pgfkeysvalueof{/tikz/ampkeys/linestyle}}%
		}
	},
	/tikz/ampkeys/drawer/.code n args={5}{%
		\tikzset{
			ampshape,
			minimum height=#2*\NODESIZE,
			minimum width=#3*\NODESIZE,
			append after command={
				\pgfextra{\let\bdr=\tikzlastnode%
				\if#1e
					\draw[draw=#4, fill=#4l, #5] (\bdr.sw) to (\bdr.nw) to (\bdr.e) to cycle {};
				\fi
				\if#1w
					\draw[draw=#4, fill=#4l, #5] (\bdr.se) to (\bdr.ne) to (\bdr.w) to cycle {};
				\fi
				\if#1n
					\draw[draw=#4, fill=#4l, #5] (\bdr.se) to (\bdr.sw) to (\bdr.n) to cycle {};
				\fi
				\if#1s
					\draw[draw=#4, fill=#4l, #5] (\bdr.ne) to (\bdr.nw) to (\bdr.s) to cycle {};
				\fi
				}
			}
		}
	},
}
\newcount\portcount

\pgfdeclareshape{aomshape}{
	\savedmacro\direction{
		\edef\direction{\pgfkeysvalueof{/tikz/aomkeys/direction}}%
	}
	\saveddimen\minwidth{
		\pgfmathsetlength\pgf@x{\pgfshapeminwidth}%
	}
	\saveddimen\minheight{
		\pgfmathsetlength\pgf@x{\pgfshapeminheight}%
	}
	\inheritsavedanchors[from=basic]

	\inheritanchor[from=basic]{center}
	\inheritanchor[from=basic]{mid}
	\inheritanchor[from=basic]{base}
	\inheritanchor[from=basic]{north}
	\inheritanchor[from=basic]{south}
	\inheritanchor[from=basic]{west}
	\inheritanchor[from=basic]{mid west}
	\inheritanchor[from=basic]{base west}
	\inheritanchor[from=basic]{north west}
	\inheritanchor[from=basic]{south west}
	\inheritanchor[from=basic]{east}
	\inheritanchor[from=basic]{mid east}
	\inheritanchor[from=basic]{base east}
	\inheritanchor[from=basic]{north east}
	\inheritanchor[from=basic]{south east}
	\inheritanchor[from=basic]{west south west}%
	\inheritanchor[from=basic]{west north west}%
	\inheritanchor[from=basic]{east north east}%
	\inheritanchor[from=basic]{east south east}%
	\inheritanchor[from=basic]{north north west}%
	\inheritanchor[from=basic]{north north east}%
	\inheritanchor[from=basic]{south south west}%
	\inheritanchor[from=basic]{south south east}%

	\inheritanchorborder[from=basic]
	\inheritbackgroundpath[from=basic]

	\pgfutil@g@addto@macro\pgf@sh@s@aomshape{%
		\pgfutil@ifundefined{pgf@anchor@aomshape@in0}{
			\expandafter\xdef\csname pgf@anchor@aomshape@in0\endcsname{%
				\noexpand\aomshape@port{0}
			}%
		}{}%
		\pgfutil@ifundefined{pgf@anchor@aomshape@in}{
			\expandafter\xdef\csname pgf@anchor@aomshape@in\endcsname{%
				\noexpand\aomshape@port{0}
			}%
		}{}%
		\pgfutil@ifundefined{pgf@anchor@aomshape@out0}{
			\expandafter\xdef\csname pgf@anchor@aomshape@out0\endcsname{%
				\noexpand\aomshape@port{1}
			}%
		}{}%
		\pgfutil@ifundefined{pgf@anchor@aomshape@out}{
			\expandafter\xdef\csname pgf@anchor@aomshape@out\endcsname{%
				\noexpand\aomshape@port{1}
			}%
		}{}%
	}
}

\def\aomshape@port#1{
	\northeast	

	\ifnum#1=0	
		\if\direction\direce
			\pgf@x=-\pgf@x
			\pgf@ya= \pgf@y
			\pgfmathsetlength{\pgf@y}{\pgf@ya-0.5*\minheight}%
		\fi
		\if\direction\direcw
			\pgf@x=\pgf@x
			\pgf@ya= \pgf@y
			\pgfmathsetlength{\pgf@y}{\pgf@ya-0.5*\minheight}%
		\fi
		\if\direction\direcn
			\pgf@y=-\pgf@y
			\pgf@xa=\pgf@x
			\pgfmathsetlength{\pgf@x}{\pgf@xa-0.5*\minwidth}%
		\fi
		\if\direction\direcs
			\pgf@y=\pgf@y
			\pgf@xa= \pgf@x
			\pgfmathsetlength{\pgf@x}{\pgf@xa-0.5*\minwidth}%
		\fi
	\else	
		\if\direction\direce
			\pgf@x=\pgf@x
			\pgf@ya= \pgf@y
			\pgfmathsetlength{\pgf@y}{\pgf@ya-0.5*\minheight}%
		\fi
		\if\direction\direcw
			\pgf@x=-\pgf@x
			\pgf@ya= \pgf@y
			\pgfmathsetlength{\pgf@y}{\pgf@ya-0.5*\minheight}%
		\fi
		\if\direction\direcn
			\pgf@y=\pgf@y
			\pgf@xa= \pgf@x
			\pgfmathsetlength{\pgf@x}{\pgf@xa-0.5*\minwidth}%
		\fi
		\if\direction\direcs
			\pgf@y=-\pgf@y
			\pgf@xa= \pgf@x
			\pgfmathsetlength{\pgf@x}{\pgf@xa-0.5*\minwidth}%
		\fi
	\fi
}

\pgfaddtoshape{aomshape}{
	\anchorlet{c}{center}
	\anchorlet{n}{north}
	\anchorlet{e}{east}
	\anchorlet{s}{south}
	\anchorlet{w}{west}
	\anchorlet{se}{south east}
	\anchorlet{sw}{south west}
	\anchorlet{ne}{north east}
	\anchorlet{nw}{north west}
	\anchorlet{wsw}{west south west}
	\anchorlet{wnw}{west north west}
	\anchorlet{ene}{east north east}
	\anchorlet{ese}{east south east}
	\anchorlet{nnw}{north north west}
	\anchorlet{nne}{north north east}
	\anchorlet{ssw}{south south west}
	\anchorlet{sse}{south south east}
}

\tikzset{
/tikz/aomkeys/.cd,
size/.initial=1,
circlesize/.initial=1,
color/.initial=O,
direction/.initial=e,
linestyle/.initial={linestyle, inner sep=0.5mm},
fillgradient/.initial=O,
/tikz/aom/.code={
\pgfqkeys{/tikz/aomkeys}{#1}%
\tikzset{/tikz/aomkeys/drawer/.expanded=%
	{\pgfkeysvalueof{/tikz/aomkeys/size}}%
	{\pgfkeysvalueof{/tikz/aomkeys/color}}%
	{\pgfkeysvalueof{/tikz/aomkeys/linestyle}}%
\if\pgfkeysvalueof{/tikz/aomkeys/direction}e
	{0}%
\fi
\if\pgfkeysvalueof{/tikz/aomkeys/direction}w
	{0}%
\fi
\if\pgfkeysvalueof{/tikz/aomkeys/direction}n
	{1}%
\fi
\if\pgfkeysvalueof{/tikz/aomkeys/direction}s
	{1}%
\fi
{\pgfkeysvalueof{/tikz/aomkeys/direction}}%
{\pgfkeysvalueof{/tikz/aomkeys/circlesize}}%
{\pgfkeysvalueof{/tikz/aomkeys/fillgradient}}%
}
},
/tikz/aomkeys/drawer/.code n args={7}{%
		\tikzset{
			aomshape,
			draw,
			minimum height = #1*\NODESIZE,
			minimum width = #1*\NODESIZE,
			#2,
			#3,
			append after command={
					\pgfextra{\let\bdr=\tikzlastnode%
						\node[#7, fit=(\bdr.nw)(\bdr.se)] (boxgradient){};

						\node[coordinate] at ($(\bdr.in)!0.25!(\bdr.out)$) (circlein){};
						\node[coordinate] at ($(\bdr.in)!0.75!(\bdr.out)$) (circleout){};

						\ifnum#4>0
							\node[coordinate] at (circleout -| \bdr.nne) (circleouttop){};
						\else
							\node[coordinate] at (circleout |- \bdr.ene) (circleouttop){};
						\fi

						\draw[---, #2, #3, fill] (\bdr.in) to (circlein) circle (0.05*#6);
						\draw[---, #2, #3, fill] (\bdr.out) to (circleout) circle (0.05*#6);

						\draw[---, #2, #3] (circlein) to (circleouttop){};

					}
				}
		}
	},
}
\newcount\portcount

\pgfdeclareshape{boxshape}{
	\savedmacro\nin{
		\edef\nin{\pgfkeysvalueof{/tikz/boxkeys/nin}}%
	}
	\savedmacro\nout{
		\edef\nout{\pgfkeysvalueof{/tikz/boxkeys/nout}}%
	}
	\savedmacro\direction{
		\edef\direction{\pgfkeysvalueof{/tikz/boxkeys/direction}}%
	}
	\inheritsavedanchors[from=basic]

	\inheritanchor[from=basic]{center}
	\inheritanchor[from=basic]{mid}
	\inheritanchor[from=basic]{base}
	\inheritanchor[from=basic]{north}
	\inheritanchor[from=basic]{south}
	\inheritanchor[from=basic]{west}
	\inheritanchor[from=basic]{mid west}
	\inheritanchor[from=basic]{base west}
	\inheritanchor[from=basic]{north west}
	\inheritanchor[from=basic]{south west}
	\inheritanchor[from=basic]{east}
	\inheritanchor[from=basic]{mid east}
	\inheritanchor[from=basic]{base east}
	\inheritanchor[from=basic]{north east}
	\inheritanchor[from=basic]{south east}
	\inheritanchor[from=basic]{west south west}%
	\inheritanchor[from=basic]{west north west}%
	\inheritanchor[from=basic]{east north east}%
	\inheritanchor[from=basic]{east south east}%
	\inheritanchor[from=basic]{north north west}%
	\inheritanchor[from=basic]{north north east}%
	\inheritanchor[from=basic]{south south west}%
	\inheritanchor[from=basic]{south south east}%

	\inheritanchorborder[from=basic]
	\inheritbackgroundpath[from=basic]

	\pgfutil@g@addto@macro\pgf@sh@s@boxshape{%
		\pgfmathsetcount{\portcount}{0}
		\pgfmathloop%
		\ifnum\the\portcount<\nin
		\pgfutil@ifundefined{pgf@anchor@boxshape@in\the\portcount}{
			\expandafter\xdef\csname pgf@anchor@boxshape@in\the\portcount\endcsname{%
				\noexpand\boxshape@port[\the\portcount]{0}
			}%
		}{}%
		\ifnum\the\portcount=0
			\pgfutil@ifundefined{pgf@anchor@boxshape@in}{%
				\expandafter\xdef\csname pgf@anchor@boxshape@in\endcsname{%
					\noexpand\boxshape@port[\the\portcount]{0}
				}%
			}{}%
		\fi
		\pgfmathaddtocount{\portcount}{1}	
		\repeatpgfmathloop
		\pgfmathsetcount{\portcount}{0}
		\pgfmathloop%
		\ifnum\the\portcount<\nout
		\pgfutil@ifundefined{pgf@anchor@boxshape@out\the\portcount}{%
			\expandafter\xdef\csname pgf@anchor@boxshape@out\the\portcount\endcsname{%
				\noexpand\boxshape@port[\the\portcount]{1}
			}%
		}{}%
		\ifnum\the\portcount=0
			\pgfutil@ifundefined{pgf@anchor@boxshape@out}{%
				\expandafter\xdef\csname pgf@anchor@boxshape@out\endcsname{%
					\noexpand\boxshape@port[\the\portcount]{1}
				}%
			}{}%
		\fi
		\pgfmathaddtocount{\portcount}{1}	
		\repeatpgfmathloop
	}
}

\def\boxshape@port[#1]#2{
	\northeast \pgf@xa=\pgf@x \pgf@ya=\pgf@y
	\southwest \pgf@xb=\pgf@x \pgf@yb=\pgf@y

	\ifnum#2=0	
		\if\direction\direce	
			\pgf@x=\pgf@xb
			\pgf@yc=\pgf@ya \advance\pgf@yc by -\pgf@yb	
			\pgfmathsetlength{\pgf@y}{\pgf@ya-(#1 + 0.5)*(\pgf@yc/\nin)}%
		\fi
		\if\direction\direcw
			\pgf@x=\pgf@xa
			\pgf@yc=\pgf@ya \advance\pgf@yc by -\pgf@yb	
			\pgfmathsetlength{\pgf@y}{\pgf@ya-(#1 + 0.5)*(\pgf@yc/\nin)}%
		\fi
		\if\direction\direcn
			\pgf@y=\pgf@yb
			\pgf@xc=\pgf@xa \advance\pgf@xc by -\pgf@xb	
			\pgfmathsetlength{\pgf@x}{\pgf@xb+(#1 + 0.5)*(\pgf@xc/\nin)}%
		\fi
		\if\direction\direcs
			\pgf@y=\pgf@ya
			\pgf@xc=\pgf@xa \advance\pgf@xc by -\pgf@xb	
			\pgfmathsetlength{\pgf@x}{\pgf@xb+(#1 + 0.5)*(\pgf@xc/\nin)}%
		\fi
	\else	
		\if\direction\direce	
			\pgf@x=\pgf@xa
			\pgf@yc=\pgf@ya \advance\pgf@yc by -\pgf@yb	
			\pgfmathsetlength{\pgf@y}{\pgf@ya-(#1 + 0.5)*(\pgf@yc/\nout)}%
		\fi
		\if\direction\direcw
			\pgf@x=\pgf@xb
			\pgf@yc=\pgf@ya \advance\pgf@yc by -\pgf@yb	
			\pgfmathsetlength{\pgf@y}{\pgf@ya-(#1 + 0.5)*(\pgf@yc/\nout)}%
		\fi
		\if\direction\direcn
			\pgf@y=\pgf@ya
			\pgf@xc=\pgf@xa \advance\pgf@xc by -\pgf@xb	
			\pgfmathsetlength{\pgf@x}{\pgf@xb+(#1 + 0.5)*(\pgf@xc/\nout)}%
		\fi
		\if\direction\direcs
			\pgf@y=\pgf@yb
			\pgf@xc=\pgf@xa \advance\pgf@xc by -\pgf@xb	
			\pgfmathsetlength{\pgf@x}{\pgf@xb+(#1 + 0.5)*(\pgf@xc/\nout)}%
		\fi
	\fi
}

\pgfaddtoshape{boxshape}{
	\anchorlet{c}{center}
	\anchorlet{n}{north}
	\anchorlet{e}{east}
	\anchorlet{s}{south}
	\anchorlet{w}{west}
	\anchorlet{se}{south east}
	\anchorlet{sw}{south west}
	\anchorlet{ne}{north east}
	\anchorlet{nw}{north west}
	\anchorlet{wsw}{west south west}
	\anchorlet{wnw}{west north west}
	\anchorlet{ene}{east north east}
	\anchorlet{ese}{east south east}
	\anchorlet{nnw}{north north west}
	\anchorlet{nne}{north north east}
	\anchorlet{ssw}{south south west}
	\anchorlet{sse}{south south east}
}

\tikzset{
/tikz/boxkeys/.cd,
height/.initial=0.5,
width/.initial=1,
color/.initial=O,
direction/.initial=e,
linestyle/.initial={linestyle, inner sep=0.5mm},
nin/.initial=1,
nout/.initial=1,
draw/.initial=1,
/tikz/box/.code={
\pgfqkeys{/tikz/boxkeys}{#1}%
\tikzset{/tikz/boxkeys/drawer/.expanded=%
\if\pgfkeysvalueof{/tikz/boxkeys/direction}e
	{\pgfkeysvalueof{/tikz/boxkeys/width}}%
	{\pgfkeysvalueof{/tikz/boxkeys/height}}%
	{0}%
	{-90}%
\fi
\if\pgfkeysvalueof{/tikz/boxkeys/direction}w
	{\pgfkeysvalueof{/tikz/boxkeys/width}}%
	{\pgfkeysvalueof{/tikz/boxkeys/height}}%
	{0}%
	{90}%
\fi
\if\pgfkeysvalueof{/tikz/boxkeys/direction}n
	{\pgfkeysvalueof{/tikz/boxkeys/width}}%
	{\pgfkeysvalueof{/tikz/boxkeys/height}}%
	{1}%
	{0}%
\fi
\if\pgfkeysvalueof{/tikz/boxkeys/direction}s
	{\pgfkeysvalueof{/tikz/boxkeys/width}}%
	{\pgfkeysvalueof{/tikz/boxkeys/height}}%
	{1}%
	{180}%
\fi
{\pgfkeysvalueof{/tikz/boxkeys/color}}%
{\pgfkeysvalueof{/tikz/boxkeys/linestyle}}%
\ifnum\pgfkeysvalueof{/tikz/boxkeys/draw}>0%
	{draw}%
\else
	{}
\fi
}
},
/tikz/boxkeys/drawer/.code n args={7}{%
		\tikzset{
			boxshape,
			#7,
			#6,
			#5,
			minimum height=
			\ifnum#3>0	
				#1*\NODESIZE
			\else
				#2*\NODESIZE
			\fi
			,minimum width=
			\ifnum#3>0
				#2*\NODESIZE
			\else
				#1*\NODESIZE
			\fi
		}
	},
}
\pgfdeclareshape{beshape}{ 
    \savedmacro\nports{
        \edef\nports{\pgfkeysvalueof{/tikz/bekeys/nports}}%
    }
    \savedmacro\direction{
        \edef\direction{\pgfkeysvalueof{/tikz/bekeys/direction}}%
    }
    \savedmacro\inverted{
        \edef\inverted{\pgfkeysvalueof{/tikz/bekeys/inverted}}%
    }
    \savedmacro\ninports{
        \ifnum\inverted=0
        \edef\ninports{\nports}
        \else
        \edef\ninports{1}%
        \fi
    }
    \savedmacro\noutports{
        \ifnum\inverted=0
        \edef\noutports{1}%
        \else
        \edef\noutports{\nports}%
        \fi
    }
    \inheritsavedanchors[from=basic]

    \inheritanchor[from=basic]{center}
    \inheritanchor[from=basic]{mid}
    \inheritanchor[from=basic]{base}
    \inheritanchor[from=basic]{north}
    \inheritanchor[from=basic]{south}
    \inheritanchor[from=basic]{west}
    \inheritanchor[from=basic]{mid west}
    \inheritanchor[from=basic]{base west}
    \inheritanchor[from=basic]{north west}
    \inheritanchor[from=basic]{south west}
    \inheritanchor[from=basic]{east}
    \inheritanchor[from=basic]{mid east}
    \inheritanchor[from=basic]{base east}
    \inheritanchor[from=basic]{north east}
    \inheritanchor[from=basic]{south east}
    \inheritanchor[from=basic]{west south west}%
    \inheritanchor[from=basic]{west north west}%
    \inheritanchor[from=basic]{east north east}%
    \inheritanchor[from=basic]{east south east}%
    \inheritanchor[from=basic]{north north west}%
    \inheritanchor[from=basic]{north north east}%
    \inheritanchor[from=basic]{south south west}%
    \inheritanchor[from=basic]{south south east}%

    \inheritanchorborder[from=basic]
    \inheritbackgroundpath[from=basic]

    \pgfutil@g@addto@macro\pgf@sh@s@beshape{%
        \pgfmathsetcount{\portcount}{0}
        \pgfmathloop%
        \ifnum\the\portcount<\nports
        \ifnum\the\portcount<\ninports
        \pgfutil@ifundefined{pgf@anchor@beshape@in\the\portcount}{
            \expandafter\xdef\csname pgf@anchor@beshape@in\the\portcount\endcsname{%
                \noexpand\beshape@port[\the\portcount]{0}
            }%
        }{}%
        \ifnum\the\portcount=0
        \pgfutil@ifundefined{pgf@anchor@beshape@in}{%
            \expandafter\xdef\csname pgf@anchor@beshape@in\endcsname{%
                \noexpand\beshape@port[\the\portcount]{0}
            }%
        }{}%
        \fi
        \fi
        \ifnum\the\portcount<\noutports
        \pgfutil@ifundefined{pgf@anchor@beshape@out\the\portcount}{%
            \expandafter\xdef\csname pgf@anchor@beshape@out\the\portcount\endcsname{%
                \noexpand\beshape@port[\the\portcount]{1}
            }%
        }{}%
        \ifnum\the\portcount=0
        \pgfutil@ifundefined{pgf@anchor@beshape@out}{%
            \expandafter\xdef\csname pgf@anchor@beshape@out\endcsname{%
                \noexpand\beshape@port[\the\portcount]{1}
            }%
        }{}%
        \fi
        \fi
        \pgfmathaddtocount{\portcount}{1}    
        \repeatpgfmathloop
    }
}

\def\beshape@port[#1]#2{
    \northeast \pgf@xa=\pgf@x \pgf@ya=\pgf@y
    \southwest \pgf@xb=\pgf@x \pgf@yb=\pgf@y

    \ifnum#2=0
    \ifnum\inverted=0
    \def\chooseports{0}
    \else
    \def\chooseports{1}
    \fi
    \else
    \ifnum\inverted=0
    \def\chooseports{1}
    \else
    \def\chooseports{0}
    \fi
    \fi

    \ifnum\chooseports=0    
    \if\direction\direce
    \pgf@x=\pgf@xb
    \pgf@yc=\pgf@ya \advance\pgf@yc by -\pgf@yb    
    \pgfmathsetlength{\pgf@y}{\pgf@ya-(#1 + 0.5)*(\pgf@yc/\nports)}%
    \fi
    \if\direction\direcw
    \pgf@x=\pgf@xa
    \pgf@yc=\pgf@ya \advance\pgf@yc by -\pgf@yb    
    \pgfmathsetlength{\pgf@y}{\pgf@ya-(#1 + 0.5)*(\pgf@yc/\nports)}%
    \fi
    \if\direction\direcn
    \pgf@y=\pgf@yb
    \pgf@xc=\pgf@xa \advance\pgf@xc by -\pgf@xb    
    \pgfmathsetlength{\pgf@x}{\pgf@xb+(#1 + 0.5)*(\pgf@xc/\nports)}%
    \fi
    \if\direction\direcs
    \pgf@y=\pgf@ya
    \pgf@xc=\pgf@xa \advance\pgf@xc by -\pgf@xb    
    \pgfmathsetlength{\pgf@x}{\pgf@xb+(#1 + 0.5)*(\pgf@xc/\nports)}%
    \fi
    \else    
    \if\direction\direce
    \pgf@x=\pgf@xa
    \pgf@yc=\pgf@ya \advance\pgf@yc by -\pgf@yb    
    \pgfmathsetlength{\pgf@y}{\pgf@ya-0.5\pgf@yc}%
    \fi
    \if\direction\direcw
    \pgf@x=\pgf@xb
    \pgf@yc=\pgf@ya \advance\pgf@yc by -\pgf@yb    
    \pgfmathsetlength{\pgf@y}{\pgf@ya-0.5\pgf@yc}%
    \fi
    \if\direction\direcn
    \pgf@y=\pgf@ya
    \pgf@xc=\pgf@xa \advance\pgf@xc by -\pgf@xb    
    \pgfmathsetlength{\pgf@x}{\pgf@xa-0.5\pgf@xc}%
    \fi
    \if\direction\direcs
    \pgf@y=\pgf@yb
    \pgf@xc=\pgf@xa \advance\pgf@xc by -\pgf@xb    
    \pgfmathsetlength{\pgf@x}{\pgf@xa-0.5\pgf@xc}%
    \fi
    \fi
}

\pgfaddtoshape{beshape}{
    \anchorlet{c}{center}
    \anchorlet{n}{north}
    \anchorlet{e}{east}
    \anchorlet{s}{south}
    \anchorlet{w}{west}
    \anchorlet{se}{south east}
    \anchorlet{sw}{south west}
    \anchorlet{ne}{north east}
    \anchorlet{nw}{north west}
    \anchorlet{wsw}{west south west}
    \anchorlet{wnw}{west north west}
    \anchorlet{ene}{east north east}
    \anchorlet{ese}{east south east}
    \anchorlet{nnw}{north north west}
    \anchorlet{nne}{north north east}
    \anchorlet{ssw}{south south west}
    \anchorlet{sse}{south south east}
}

\tikzset{
    /tikz/bekeys/.cd,
    height/.initial=1,
    width/.initial=0.5,
    color/.initial=O,
    direction/.initial=e,
    linestyle/.initial={linestyle, rounded corners = 0},
    nports/.initial=3,
    inverted/.initial=0,    
    sbe/.initial=0,
    angle/.initial=60,
    /tikz/be/.code={
        \pgfqkeys{/tikz/bekeys}{#1}%
        \tikzset{/tikz/bekeys/drawer/.expanded=%
            \if\pgfkeysvalueof{/tikz/bekeys/direction}e
                {\pgfkeysvalueof{/tikz/bekeys/width}}%
                {\pgfkeysvalueof{/tikz/bekeys/height}}%
                {0}%
                {-90}%
            \fi
            \if\pgfkeysvalueof{/tikz/bekeys/direction}w
                {\pgfkeysvalueof{/tikz/bekeys/width}}%
                {\pgfkeysvalueof{/tikz/bekeys/height}}%
                {0}%
                {90}%
            \fi
            \if\pgfkeysvalueof{/tikz/bekeys/direction}n
                {\pgfkeysvalueof{/tikz/bekeys/width}}%
                {\pgfkeysvalueof{/tikz/bekeys/height}}%
                {1}%
                {0}%
            \fi
            \if\pgfkeysvalueof{/tikz/bekeys/direction}s
                {\pgfkeysvalueof{/tikz/bekeys/width}}%
                {\pgfkeysvalueof{/tikz/bekeys/height}}%
                {1}%
                {180}%
            \fi
            {\pgfkeysvalueof{/tikz/bekeys/color}}%
            {\pgfkeysvalueof{/tikz/bekeys/linestyle}}%
            {\pgfkeysvalueof{/tikz/bekeys/sbe}}%
            {\pgfkeysvalueof{/tikz/bekeys/angle}}%
        }
    },
    /tikz/bekeys/drawer/.code n args={8}{%
        \tikzset{
            beshape,
            #6,
            #5,
            minimum height=
            \ifnum#3>0    
            #1*\NODESIZE
            \else
            #2*\NODESIZE
            \fi
            ,minimum width=
            \ifnum#3>0
            #2*\NODESIZE
            \else
            #1*\NODESIZE
            \fi
            ,append after command={
                \pgfextra{
                    \let\bdr=\tikzlastnode%
                    \node[trapezium, line width = \NODETHICKNESS, minimum height=#1*\NODESIZE, minimum width=#2*\NODESIZE, trapezium stretches=true, rotate=#4, trapezium angle=70, inner sep=0.001mm, #5, #6] at (\bdr) (trap) {};

                    \node[rectangle, line width = \NODETHICKNESS, minimum height=#1*\NODESIZE, minimum width=#2*\NODESIZE, anchor=north, rotate=#4, #5, #6] at (trap.south) (r1) {};

                    \tikzmath{coordinate \C;
                    \C = (trap.top left corner)-(trap.top right corner);
                    \distAB = sqrt((\Cx)^2+(\Cy)^2);
                    }

                    \node[rectangle, line width = \NODETHICKNESS, minimum height=#1*\NODESIZE, minimum width=\distAB, anchor=south, rotate=#4, #5, #6, red] at (trap.north) (r2) {};

					\draw[#5, #6] (trap.top left corner) to (r2.north west) to (r2.north east) to (trap.top right corner) to (trap.bottom right corner) to (r1.south east) to (r1.south west) to (r1.north west) to cycle;

                }
            }
        }
    },
}
\pgfdeclareshape{couplershape}{	

	\inheritsavedanchors[from=basic] 

	\inheritanchor[from=basic]{center}
	\inheritanchor[from=basic]{mid}		
	\inheritanchor[from=basic]{base}	
	\inheritanchor[from=basic]{north}
	\inheritanchor[from=basic]{south}		
	\inheritanchor[from=basic]{west}		
	\inheritanchor[from=basic]{mid west}				
	\inheritanchor[from=basic]{base west}		
	\inheritanchor[from=basic]{north west}		
	\inheritanchor[from=basic]{south west}		
	\inheritanchor[from=basic]{east}
	\inheritanchor[from=basic]{mid east}
	\inheritanchor[from=basic]{base east}	
	\inheritanchor[from=basic]{north east}
	\inheritanchor[from=basic]{south east}
	\inheritanchor[from=basic]{west south west}%
	\inheritanchor[from=basic]{west north west}%
	\inheritanchor[from=basic]{east north east}%
	\inheritanchor[from=basic]{east south east}%
	\inheritanchor[from=basic]{north north west}%
	\inheritanchor[from=basic]{north north east}%
	\inheritanchor[from=basic]{south south west}%
	\inheritanchor[from=basic]{south south east}%
	
	\inheritanchorborder[from=basic]	
	\inheritbackgroundpath[from=basic]
}

\pgfaddtoshape{couplershape}{
	\anchorlet{in}{center}		
	\anchorlet{in0}{center}
	\anchorlet{out}{center}
	\anchorlet{out0}{center}
	\anchorlet{c}{center}		
	\anchorlet{n}{north}
	\anchorlet{e}{east}
	\anchorlet{s}{south}
	\anchorlet{w}{west}			
	\anchorlet{se}{south east}
	\anchorlet{sw}{south west}
	\anchorlet{ne}{north east}
	\anchorlet{nw}{north west}
	\anchorlet{wsw}{west south west}
	\anchorlet{wnw}{west north west}
	\anchorlet{ene}{east north east}
	\anchorlet{ese}{east south east}
	\anchorlet{nnw}{north north west}
	\anchorlet{nne}{north north east}
	\anchorlet{ssw}{south south west}
	\anchorlet{sse}{south south east}
}

\tikzset{
	/tikz/couplerkeys/.cd,
	size/.initial=0.2,
	color/.initial=O,
	rotation/.initial=0,
	heightwidthratio/.initial=0.5,
	/tikz/coupler/.code={
		\pgfqkeys{/tikz/couplerkeys}{#1}%
		\tikzset{/tikz/couplerkeys/drawer/.expanded=%
			{\pgfkeysvalueof{/tikz/couplerkeys/size}}%
			{\pgfkeysvalueof{/tikz/couplerkeys/color}}%
			{\pgfkeysvalueof{/tikz/couplerkeys/rotation}}%
			{\pgfkeysvalueof{/tikz/couplerkeys/heightwidthratio}}%
		}
	},
	/tikz/couplerkeys/drawer/.code n args={4}{%
		\tikzset{
			couplershape,
			minimum height=#1*\NODESIZE
			\ifnum#3<1
				\ifnum#3>-1
					*#4
				\fi
			\fi
			,minimum width=#1*\NODESIZE
			\ifnum#3<91
				\ifnum#3>89
					*#4
				\fi
			\fi
			\ifnum#3<-89
				\ifnum#3>-91
					*#4
				\fi
			\fi
			,#2,
			append after command={
				\pgfextra{\let\bdr=\tikzlastnode%
				\node[ellipse, fill, #2, rotate=#3, outer sep = 0, minimum width=#1*\NODESIZE, minimum height=#1*#4*\NODESIZE] at (\bdr.center){};
				}
			}
		}
	},
}
\pgfdeclareshape{fibershape}{
	\savedmacro\direction{
		\edef\direction{\pgfkeysvalueof{/tikz/fiberkeys/direction}}%
	}
	\savedmacro\flip{
		\edef\flip{\pgfkeysvalueof{/tikz/fiberkeys/flip}}%
	}
	\saveddimen\minwidth{
		\pgfmathsetlength\pgf@x{\pgfshapeminwidth}%
	}
	\saveddimen\minheight{
		\pgfmathsetlength\pgf@x{\pgfshapeminheight}%
	}
	\inheritsavedanchors[from=basic] 
	
	\inheritanchor[from=basic]{center}
	\inheritanchor[from=basic]{mid}		
	\inheritanchor[from=basic]{base}	
	\inheritanchor[from=basic]{north}
	\inheritanchor[from=basic]{south}		
	\inheritanchor[from=basic]{west}		
	\inheritanchor[from=basic]{mid west}				
	\inheritanchor[from=basic]{base west}		
	\inheritanchor[from=basic]{north west}		
	\inheritanchor[from=basic]{south west}		
	\inheritanchor[from=basic]{east}
	\inheritanchor[from=basic]{mid east}
	\inheritanchor[from=basic]{base east}	
	\inheritanchor[from=basic]{north east}
	\inheritanchor[from=basic]{south east}
	\inheritanchor[from=basic]{west south west}%
	\inheritanchor[from=basic]{west north west}%
	\inheritanchor[from=basic]{east north east}%
	\inheritanchor[from=basic]{east south east}%
	\inheritanchor[from=basic]{north north west}%
	\inheritanchor[from=basic]{north north east}%
	\inheritanchor[from=basic]{south south west}%
	\inheritanchor[from=basic]{south south east}%
	
	\inheritanchorborder[from=basic]	
	\inheritbackgroundpath[from=basic]

    \pgfutil@g@addto@macro\pgf@sh@s@fibershape{%
        \pgfutil@ifundefined{pgf@anchor@fibershape@in0}{
	        \expandafter\xdef\csname pgf@anchor@fibershape@in0\endcsname{%
	            \noexpand\fibershape@port{0}
	        }%
	    }{}%
        \pgfutil@ifundefined{pgf@anchor@fibershape@in}{
	        \expandafter\xdef\csname pgf@anchor@fibershape@in\endcsname{%
	            \noexpand\fibershape@port{0}
	        }%
	    }{}%
        \pgfutil@ifundefined{pgf@anchor@fibershape@out0}{
	        \expandafter\xdef\csname pgf@anchor@fibershape@out0\endcsname{%
	            \noexpand\fibershape@port{1}
	        }%
	    }{}%
        \pgfutil@ifundefined{pgf@anchor@fibershape@out}{
	        \expandafter\xdef\csname pgf@anchor@fibershape@out\endcsname{%
	            \noexpand\fibershape@port{1}
	        }%
	    }{}%
	}
}

\def\fibershape@port#1{
    \northeast	

    \ifnum#1=0	
	    \if\direction\direce
			\pgf@x=-\pgf@x
	    	\if\flip\flipfalse
		    	\pgf@y=-\pgf@y
		    \else
		    	\pgf@y=\pgf@y
		    \fi
		\fi
	    \if\direction\direcw
			\pgf@x=\pgf@x
	    	\if\flip\flipfalse
		    	\pgf@y=-\pgf@y
		    \else
		    	\pgf@y=\pgf@y
		    \fi
		\fi
	    \if\direction\direcn
			\pgf@y=-\pgf@y
	    	\if\flip\flipfalse
		    	\pgf@x=-\pgf@x
		    \else
		    	\pgf@x=\pgf@x
		    \fi
		\fi
	    \if\direction\direcs
			\pgf@y=\pgf@y
	    	\if\flip\flipfalse
		    	\pgf@x=-\pgf@x
		    \else
		    	\pgf@x=\pgf@x
		    \fi
		\fi
	\else	
	    \if\direction\direce
			\pgf@x=\pgf@x
	    	\if\flip\flipfalse
		    	\pgf@y=-\pgf@y
		    \else
		    	\pgf@y=\pgf@y
		    \fi
		\fi
	    \if\direction\direcw
			\pgf@x=-\pgf@x
	    	\if\flip\flipfalse
		    	\pgf@y=-\pgf@y
		    \else
		    	\pgf@y=\pgf@y
		    \fi
		\fi
	    \if\direction\direcn
			\pgf@y=\pgf@y
	    	\if\flip\flipfalse
		    	\pgf@x=-\pgf@x
		    \else
		    	\pgf@x=\pgf@x
		    \fi
		\fi
	    \if\direction\direcs
			\pgf@y=-\pgf@y
	    	\if\flip\flipfalse
		    	\pgf@x=-\pgf@x
		    \else
		    	\pgf@x=\pgf@x
		    \fi
		\fi
	\fi
}

\pgfaddtoshape{fibershape}{
	\anchorlet{c}{center}		
	\anchorlet{n}{north}
	\anchorlet{e}{east}
	\anchorlet{s}{south}
	\anchorlet{w}{west}			
	\anchorlet{se}{south east}
	\anchorlet{sw}{south west}
	\anchorlet{ne}{north east}
	\anchorlet{nw}{north west}
	\anchorlet{wsw}{west south west}
	\anchorlet{wnw}{west north west}
	\anchorlet{ene}{east north east}
	\anchorlet{ese}{east south east}
	\anchorlet{nnw}{north north west}
	\anchorlet{nne}{north north east}
	\anchorlet{ssw}{south south west}
	\anchorlet{sse}{south south east}
}

\tikzset{
	/tikz/fiberkeys/.cd,
	size/.initial=1,
	color/.initial=C0,
	direction/.initial=e,
	linestyle/.initial={linestyle},
	flip/.initial={0},
	drawbase/.initial={1},
	/tikz/fiber/.code={
		\pgfqkeys{/tikz/fiberkeys}{#1}%
		\tikzset{/tikz/fiberkeys/drawer/.expanded=%
			{\pgfkeysvalueof{/tikz/fiberkeys/direction}}%
			{\pgfkeysvalueof{/tikz/fiberkeys/size}}%
			{\pgfkeysvalueof{/tikz/fiberkeys/color}}%
			{\pgfkeysvalueof{/tikz/fiberkeys/linestyle}}%
			\if\pgfkeysvalueof{/tikz/fiberkeys/direction}e
				{a}%
				\if\pgfkeysvalueof{/tikz/fiberkeys/flip}0
					{south}%
				\else
					{north}%
				\fi
				{\pgfkeysvalueof{/tikz/fiberkeys/size}}
				{\pgfkeysvalueof{/tikz/fiberkeys/size} * 0.5}
			\fi
			\if\pgfkeysvalueof{/tikz/fiberkeys/direction}w
				{a}%
				\if\pgfkeysvalueof{/tikz/fiberkeys/flip}0
					{south}%
				\else
					{north}%
				\fi
				{\pgfkeysvalueof{/tikz/fiberkeys/size}}
				{\pgfkeysvalueof{/tikz/fiberkeys/size} * 0.5}
			\fi
			\if\pgfkeysvalueof{/tikz/fiberkeys/direction}n
				{b}%
				\if\pgfkeysvalueof{/tikz/fiberkeys/flip}0
					{west}%
				\else
					{east}%
				\fi
				{\pgfkeysvalueof{/tikz/fiberkeys/size} * 0.5}
				{\pgfkeysvalueof{/tikz/fiberkeys/size}}
			\fi
			\if\pgfkeysvalueof{/tikz/fiberkeys/direction}s
				{b}%
				\if\pgfkeysvalueof{/tikz/fiberkeys/flip}0
					{west}%
				\else
					{east}%
				\fi
				{\pgfkeysvalueof{/tikz/fiberkeys/size} * 0.5}
				{\pgfkeysvalueof{/tikz/fiberkeys/size}}
			\fi
			{\pgfkeysvalueof{/tikz/fiberkeys/drawbase}}%
		}
	},
	/tikz/fiberkeys/drawer/.code n args={9}{%
		\tikzset{
			fibershape,
			minimum width=#7*\NODESIZE,
			minimum height=#8*\NODESIZE,
			append after command={
				\pgfextra{\let\bdr=\tikzlastnode%
				\if#5a	
					\ifnum#9>0
						\draw[#3, #4] (\bdr.#6 west) to (\bdr.#6 east) {};
					\fi
					\node[draw=#3, #4, circle, minimum size=#2*0.5*\NODESIZE, anchor=#6] at ([xshift=-0.1*#2*\NODESIZE]\bdr.#6) () {};
					\node[draw=#3, #4, circle, minimum size=#2*0.5*\NODESIZE, anchor=#6] at (\bdr.#6) () {};
					\node[draw=#3, #4, circle, minimum size=#2*0.5*\NODESIZE, anchor=#6] at ([xshift=0.1*#2*\NODESIZE]\bdr.#6) () {};
				\fi
				\if#5b	
					\ifnum#9>0
						\draw[#3, #4] (\bdr.north #6) to (\bdr.south #6) {};
					\fi
					\node[draw=#3, #4, circle, minimum size=#2*0.5*\NODESIZE, anchor=#6] at ([yshift=0.1*#2*\NODESIZE]\bdr.#6) () {};
					\node[draw=#3, #4, circle, minimum size=#2*0.5*\NODESIZE, anchor=#6] at (\bdr.#6) () {};
					\node[draw=#3, #4, circle, minimum size=#2*0.5*\NODESIZE, anchor=#6] at ([yshift=-0.1*#2*\NODESIZE]\bdr.#6) () {};
				\fi
				}
			}
		}
	},
}
\newcount\portcount

\pgfdeclareshape{fiberswitchshape}{
	\savedmacro\nin{
		\edef\nin{\pgfkeysvalueof{/tikz/fiberswitchkeys/nin}}%
	}
	\savedmacro\nout{
		\edef\nout{\pgfkeysvalueof{/tikz/fiberswitchkeys/nout}}%
	}
	\savedmacro\direction{
		\edef\direction{\pgfkeysvalueof{/tikz/fiberswitchkeys/direction}}%
	}
	\inheritsavedanchors[from=basic] 
	
	\inheritanchor[from=basic]{center}
	\inheritanchor[from=basic]{mid}		
	\inheritanchor[from=basic]{base}	
	\inheritanchor[from=basic]{north}
	\inheritanchor[from=basic]{south}		
	\inheritanchor[from=basic]{west}		
	\inheritanchor[from=basic]{mid west}				
	\inheritanchor[from=basic]{base west}		
	\inheritanchor[from=basic]{north west}		
	\inheritanchor[from=basic]{south west}		
	\inheritanchor[from=basic]{east}
	\inheritanchor[from=basic]{mid east}
	\inheritanchor[from=basic]{base east}	
	\inheritanchor[from=basic]{north east}
	\inheritanchor[from=basic]{south east}
	\inheritanchor[from=basic]{west south west}%
	\inheritanchor[from=basic]{west north west}%
	\inheritanchor[from=basic]{east north east}%
	\inheritanchor[from=basic]{east south east}%
	\inheritanchor[from=basic]{north north west}%
	\inheritanchor[from=basic]{north north east}%
	\inheritanchor[from=basic]{south south west}%
	\inheritanchor[from=basic]{south south east}%
	
	\inheritanchorborder[from=basic]	
	\inheritbackgroundpath[from=basic]

    \pgfutil@g@addto@macro\pgf@sh@s@fiberswitchshape{%
        \pgfmathsetcount{\portcount}{0}
        \pgfmathloop%
        \ifnum\the\portcount<\nin
	        \pgfutil@ifundefined{pgf@anchor@fiberswitchshape@in\the\portcount}{
		        \expandafter\xdef\csname pgf@anchor@fiberswitchshape@in\the\portcount\endcsname{%
		            \noexpand\fiberswitchshape@port[\the\portcount]{0}
		        }%
		    }{}%
	        \ifnum\the\portcount=0
    		    \pgfutil@ifundefined{pgf@anchor@fiberswitchshape@in}{%
		        \expandafter\xdef\csname pgf@anchor@fiberswitchshape@in\endcsname{%
		            \noexpand\fiberswitchshape@port[\the\portcount]{0}
		        }%
		        }{}%
		    \fi
	        \pgfmathaddtocount{\portcount}{1}	
	        \repeatpgfmathloop
        \pgfmathsetcount{\portcount}{0}
        \pgfmathloop%
    	\ifnum\the\portcount<\nout
	        \pgfutil@ifundefined{pgf@anchor@fiberswitchshape@out\the\portcount}{%
		        \expandafter\xdef\csname pgf@anchor@fiberswitchshape@out\the\portcount\endcsname{%
		            \noexpand\fiberswitchshape@port[\the\portcount]{1}
		        }%
		    }{}%
	        \ifnum\the\portcount=0
    		    \pgfutil@ifundefined{pgf@anchor@fiberswitchshape@out}{%
		        \expandafter\xdef\csname pgf@anchor@fiberswitchshape@out\endcsname{%
		            \noexpand\fiberswitchshape@port[\the\portcount]{1}
		        }%
		        }{}%
		    \fi
	        \pgfmathaddtocount{\portcount}{1}	
	        \repeatpgfmathloop
	}
}

\def\fiberswitchshape@port[#1]#2{
    \northeast \pgf@xa=\pgf@x \pgf@ya=\pgf@y
    \southwest \pgf@xb=\pgf@x \pgf@yb=\pgf@y
    
    \ifnum#2=0	
	    \if\direction\direce	
	    	\pgf@x=\pgf@xb
		    \pgf@yc=\pgf@ya \advance\pgf@yc by -\pgf@yb	
		    \pgfmathsetlength{\pgf@y}{\pgf@ya-(#1 + 0.5)*(\pgf@yc/\nin)}%
	    \fi
	    \if\direction\direcw
	    	\pgf@x=\pgf@xa
		    \pgf@yc=\pgf@ya \advance\pgf@yc by -\pgf@yb	
		    \pgfmathsetlength{\pgf@y}{\pgf@ya-(#1 + 0.5)*(\pgf@yc/\nin)}%
	    \fi
	    \if\direction\direcn
	    	\pgf@y=\pgf@yb
		    \pgf@xc=\pgf@xa \advance\pgf@xc by -\pgf@xb	
		    \pgfmathsetlength{\pgf@x}{\pgf@xb+(#1 + 0.5)*(\pgf@xc/\nin)}%
	    \fi
	    \if\direction\direcs
	    	\pgf@y=\pgf@ya
		    \pgf@xc=\pgf@xa \advance\pgf@xc by -\pgf@xb	
		    \pgfmathsetlength{\pgf@x}{\pgf@xb+(#1 + 0.5)*(\pgf@xc/\nin)}%
	    \fi
	\else	
	    \if\direction\direce	
	    	\pgf@x=\pgf@xa
		    \pgf@yc=\pgf@ya \advance\pgf@yc by -\pgf@yb	
		    \pgfmathsetlength{\pgf@y}{\pgf@ya-(#1 + 0.5)*(\pgf@yc/\nout)}%
	    \fi
	    \if\direction\direcw
	    	\pgf@x=\pgf@xb
		    \pgf@yc=\pgf@ya \advance\pgf@yc by -\pgf@yb	
		    \pgfmathsetlength{\pgf@y}{\pgf@ya-(#1 + 0.5)*(\pgf@yc/\nout)}%
	    \fi
	    \if\direction\direcn
	    	\pgf@y=\pgf@ya
		    \pgf@xc=\pgf@xa \advance\pgf@xc by -\pgf@xb	
		    \pgfmathsetlength{\pgf@x}{\pgf@xb+(#1 + 0.5)*(\pgf@xc/\nout)}%
	    \fi
	    \if\direction\direcs
	    	\pgf@y=\pgf@yb
		    \pgf@xc=\pgf@xa \advance\pgf@xc by -\pgf@xb	
		    \pgfmathsetlength{\pgf@x}{\pgf@xb+(#1 + 0.5)*(\pgf@xc/\nout)}%
	    \fi
	\fi
}

\pgfaddtoshape{fiberswitchshape}{
	\anchorlet{c}{center}		
	\anchorlet{n}{north}
	\anchorlet{e}{east}
	\anchorlet{s}{south}
	\anchorlet{w}{west}			
	\anchorlet{se}{south east}
	\anchorlet{sw}{south west}
	\anchorlet{ne}{north east}
	\anchorlet{nw}{north west}
	\anchorlet{wsw}{west south west}
	\anchorlet{wnw}{west north west}
	\anchorlet{ene}{east north east}
	\anchorlet{ese}{east south east}
	\anchorlet{nnw}{north north west}
	\anchorlet{nne}{north north east}
	\anchorlet{ssw}{south south west}
	\anchorlet{sse}{south south east}
}

\tikzset{
	/tikz/fiberswitchkeys/.cd,
	size/.initial=1,
	color/.initial=O,
	direction/.initial=e,
	linestyle/.initial={linestyle, inner sep=0.5mm},
	nin/.initial=1,	
	nout/.initial=3,
	/tikz/fiberswitch/.code={
		\pgfqkeys{/tikz/fiberswitchkeys}{#1}%
		\tikzset{/tikz/fiberswitchkeys/drawer/.expanded=%
			{\pgfkeysvalueof{/tikz/fiberswitchkeys/size}}%
			{\pgfkeysvalueof{/tikz/fiberswitchkeys/color}}%
			{\pgfkeysvalueof{/tikz/fiberswitchkeys/linestyle}}%
			{\pgfkeysvalueof{/tikz/fiberswitchkeys/nout}}%
			\if\pgfkeysvalueof{/tikz/fiberswitchkeys/direction}e
				{0}%
			\fi
			\if\pgfkeysvalueof{/tikz/fiberswitchkeys/direction}w
				{0}%
			\fi
			\if\pgfkeysvalueof{/tikz/fiberswitchkeys/direction}n
				{1}%
			\fi
			\if\pgfkeysvalueof{/tikz/fiberswitchkeys/direction}s
				{1}%
			\fi
			{\pgfkeysvalueof{/tikz/fiberswitchkeys/direction}}%
		}
	},
	/tikz/fiberswitchkeys/drawer/.code n args={6}{%
		\tikzset{
			fiberswitchshape,
			draw,
			minimum height = #1*\NODESIZE,
			minimum width = #1*\NODESIZE,
			#2,
			#3,
			append after command={
				\pgfextra{\let\bdr=\tikzlastnode%
						\ifnum#5>0
							\node[coordinate] at ($(\bdr.in)!0.25!(\bdr.out0 -| \bdr.in)$) (circlein){};
							\foreach \n [evaluate=\n as \nport using int(\n-1)] in {1,...,#4}{
								\node[coordinate] at ($(\bdr.out\nport)!0.25!(\bdr.in -| \bdr.out\nport)$) (circleout\nport){};
							}
						\else
							\node[coordinate] at ($(\bdr.in)!0.25!(\bdr.out0 |- \bdr.in)$) (circlein){};
							\foreach \n [evaluate=\n as \nport using int(\n-1)] in {1,...,#4}{
								\node[coordinate] at ($(\bdr.out\nport)!0.25!(\bdr.in |- \bdr.out\nport)$) (circleout\nport){};
							}
						\fi

						\draw[---, #2, #3, fill] (\bdr.in) to (circlein) circle (0.05);
						\foreach \n [evaluate=\n as \nport using int(\n-1)] in {1,...,#4}{
							\draw[---, #2, #3, fill] (\bdr.out\nport) to (circleout\nport) circle (0.05);
						}

						\tikzmath{
							int \nportmax;
							\nportmax = int(#4-1);
						}
						\draw[---, #2, #3] (circlein) to (circleout0){};
						\if#6e
							\draw[-->, #2, #3, looseness=0.8] ($(circlein)!0.6!(circleout0)$) to [out=-60, in=60]($(circlein)!0.6!(circleout\nportmax)$) {};
						\fi
						\if#6w
							\draw[-->, #2, #3, looseness=0.8] ($(circlein)!0.6!(circleout0)$) to [out=-120, in=120]($(circlein)!0.6!(circleout\nportmax)$) {};
						\fi
						\if#6s
							\draw[-->, #2, #3, looseness=0.8] ($(circlein)!0.6!(circleout0)$) to [out=-30, in=-150]($(circlein)!0.6!(circleout\nportmax)$) {};
						\fi
						\if#6n
							\draw[-->, #2, #3, looseness=0.8] ($(circlein)!0.6!(circleout0)$) to [out=30, in=150]($(circlein)!0.6!(circleout\nportmax)$) {};
						\fi

				}
			}
		}
	},
}
\pgfdeclareshape{filtershape}{
	\savedmacro\direction{
		\edef\direction{\pgfkeysvalueof{/tikz/filterkeys/direction}}%
	}
	\saveddimen\minwidth{
		\pgfmathsetlength\pgf@x{\pgfshapeminwidth}%
	}
	\saveddimen\minheight{
		\pgfmathsetlength\pgf@x{\pgfshapeminheight}%
	}
	\inheritsavedanchors[from=basic]

	\inheritanchor[from=basic]{center}
	\inheritanchor[from=basic]{mid}
	\inheritanchor[from=basic]{base}
	\inheritanchor[from=basic]{north}
	\inheritanchor[from=basic]{south}
	\inheritanchor[from=basic]{west}
	\inheritanchor[from=basic]{mid west}
	\inheritanchor[from=basic]{base west}
	\inheritanchor[from=basic]{north west}
	\inheritanchor[from=basic]{south west}
	\inheritanchor[from=basic]{east}
	\inheritanchor[from=basic]{mid east}
	\inheritanchor[from=basic]{base east}
	\inheritanchor[from=basic]{north east}
	\inheritanchor[from=basic]{south east}
	\inheritanchor[from=basic]{west south west}%
	\inheritanchor[from=basic]{west north west}%
	\inheritanchor[from=basic]{east north east}%
	\inheritanchor[from=basic]{east south east}%
	\inheritanchor[from=basic]{north north west}%
	\inheritanchor[from=basic]{north north east}%
	\inheritanchor[from=basic]{south south west}%
	\inheritanchor[from=basic]{south south east}%

	\inheritanchorborder[from=basic]
	\inheritbackgroundpath[from=basic]

	\pgfutil@g@addto@macro\pgf@sh@s@filtershape{%
		\pgfutil@ifundefined{pgf@anchor@filtershape@in0}{
			\expandafter\xdef\csname pgf@anchor@filtershape@in0\endcsname{%
				\noexpand\filtershape@port{0}
			}%
		}{}%
		\pgfutil@ifundefined{pgf@anchor@filtershape@in}{
			\expandafter\xdef\csname pgf@anchor@filtershape@in\endcsname{%
				\noexpand\filtershape@port{0}
			}%
		}{}%
		\pgfutil@ifundefined{pgf@anchor@filtershape@out0}{
			\expandafter\xdef\csname pgf@anchor@filtershape@out0\endcsname{%
				\noexpand\filtershape@port{1}
			}%
		}{}%
		\pgfutil@ifundefined{pgf@anchor@filtershape@out}{
			\expandafter\xdef\csname pgf@anchor@filtershape@out\endcsname{%
				\noexpand\filtershape@port{1}
			}%
		}{}%
	}
}

\def\filtershape@port#1{
	\northeast	

	\ifnum#1=0	
		\if\direction\direce
			\pgf@x=-\pgf@x
			\pgf@ya= \pgf@y
			\pgfmathsetlength{\pgf@y}{\pgf@ya-0.5*\minheight}%
		\fi
		\if\direction\direcw
			\pgf@x=\pgf@x
			\pgf@ya= \pgf@y
			\pgfmathsetlength{\pgf@y}{\pgf@ya-0.5*\minheight}%
		\fi
		\if\direction\direcn
			\pgf@y=-\pgf@y
			\pgf@xa=\pgf@x
			\pgfmathsetlength{\pgf@x}{\pgf@xa-0.5*\minwidth}%
		\fi
		\if\direction\direcs
			\pgf@y=\pgf@y
			\pgf@xa= \pgf@x
			\pgfmathsetlength{\pgf@x}{\pgf@xa-0.5*\minwidth}%
		\fi
	\else	
		\if\direction\direce
			\pgf@x=\pgf@x
			\pgf@ya= \pgf@y
			\pgfmathsetlength{\pgf@y}{\pgf@ya-0.5*\minheight}%
		\fi
		\if\direction\direcw
			\pgf@x=-\pgf@x
			\pgf@ya= \pgf@y
			\pgfmathsetlength{\pgf@y}{\pgf@ya-0.5*\minheight}%
		\fi
		\if\direction\direcn
			\pgf@y=\pgf@y
			\pgf@xa= \pgf@x
			\pgfmathsetlength{\pgf@x}{\pgf@xa-0.5*\minwidth}%
		\fi
		\if\direction\direcs
			\pgf@y=-\pgf@y
			\pgf@xa= \pgf@x
			\pgfmathsetlength{\pgf@x}{\pgf@xa-0.5*\minwidth}%
		\fi
	\fi
}

\pgfaddtoshape{filtershape}{
	\anchorlet{c}{center}
	\anchorlet{n}{north}
	\anchorlet{e}{east}
	\anchorlet{s}{south}
	\anchorlet{w}{west}
	\anchorlet{se}{south east}
	\anchorlet{sw}{south west}
	\anchorlet{ne}{north east}
	\anchorlet{nw}{north west}
	\anchorlet{wsw}{west south west}
	\anchorlet{wnw}{west north west}
	\anchorlet{ene}{east north east}
	\anchorlet{ese}{east south east}
	\anchorlet{nnw}{north north west}
	\anchorlet{nne}{north north east}
	\anchorlet{ssw}{south south west}
	\anchorlet{sse}{south south east}
}

\pgfmathsetmacro{\WSSSINEHEIGHT}{0.06}

\tikzset{
	/tikz/filterkeys/.cd,
	size/.initial=0.5,
	color/.initial=O,
	direction/.initial=e,
	linestyle/.initial={linestyle},
	fillgradient/.initial=O,
	/tikz/filter/.code={
			\pgfqkeys{/tikz/filterkeys}{#1}%
			\tikzset{/tikz/filterkeys/drawer/.expanded=%
					{\pgfkeysvalueof{/tikz/filterkeys/direction}}%
					{\pgfkeysvalueof{/tikz/filterkeys/size}}%
					{\pgfkeysvalueof{/tikz/filterkeys/color}}%
					{\pgfkeysvalueof{/tikz/filterkeys/linestyle}}%
					{\pgfkeysvalueof{/tikz/filterkeys/fillgradient}}%
			}
		},
	/tikz/filterkeys/drawer/.code n args={5}{%
			\tikzset{
				filtershape,
				minimum height=#2*\NODESIZE,
				minimum width=#2*\NODESIZE,
				#3,
				#4,
				draw,
				append after command={
						\pgfextra{\let\bdr=\tikzlastnode%
							\node[#5, fit=(\bdr.nw)(\bdr.se)] (boxgradient){};

							\node[coordinate] at (\bdr.wnw -| \bdr.nnw) (hl){};
							\node[coordinate] at (\bdr.ene -| \bdr.nne) (hr){};
							\draw[#3,---, rounded corners = 0] (hl) sin ($(hl)!0.25!(hr) + (0,0.002*#2*\NODESIZE)$) cos ($(hl)!0.5!(hr)$) sin ($(hl)!0.75!(hr) + (0,-0.002*#2*\NODESIZE)$) cos (hr);

							\node[coordinate] at (\bdr.w -| \bdr.nnw) (ml){};
							\node[coordinate] at (\bdr.e -| \bdr.nne) (mr){};
							\draw[#3,---, rounded corners = 0] (ml) sin ($(ml)!0.25!(mr) + (0,0.002*#2*\NODESIZE)$) cos ($(ml)!0.5!(mr)$) sin ($(ml)!0.75!(mr) + (0,-0.002*#2*\NODESIZE)$) cos (mr);

							\node[coordinate] at (\bdr.wsw -| \bdr.nnw) (ll){};
							\node[coordinate] at (\bdr.ese -| \bdr.nne) (lr){};
							\draw[#3,---, rounded corners = 0] (ll) sin ($(ll)!0.25!(lr) + (0,0.002*#2*\NODESIZE)$) cos ($(ll)!0.5!(lr)$) sin ($(ll)!0.75!(lr) + (0,-0.002*#2*\NODESIZE)$) cos (lr);

							\draw[#3, ---] ($(hl)!0.25!(hr) - (0,0.002*#2*\NODESIZE)$) -- ($(hl)!0.75!(hr) + (0,0.002*#2*\NODESIZE)$);
							\draw[#3, ---] ($(ll)!0.25!(lr) - (0,0.002*#2*\NODESIZE)$) -- ($(ll)!0.75!(lr) + (0,0.002*#2*\NODESIZE)$);
						}
					}
			}
		},
}
\newcount\portcount

\pgfdeclareshape{lensshape}{
	\savedmacro\nport{
		\edef\nport{\pgfkeysvalueof{/tikz/lenskeys/nport}}%
	}
	\inheritsavedanchors[from=basic] 
	
	\inheritanchor[from=basic]{center}
	\inheritanchor[from=basic]{mid}		
	\inheritanchor[from=basic]{base}	
	\inheritanchor[from=basic]{north}
	\inheritanchor[from=basic]{south}		
	\inheritanchor[from=basic]{west}		
	\inheritanchor[from=basic]{mid west}				
	\inheritanchor[from=basic]{base west}		
	\inheritanchor[from=basic]{north west}		
	\inheritanchor[from=basic]{south west}		
	\inheritanchor[from=basic]{east}
	\inheritanchor[from=basic]{mid east}
	\inheritanchor[from=basic]{base east}	
	\inheritanchor[from=basic]{north east}
	\inheritanchor[from=basic]{south east}
	\inheritanchor[from=basic]{west south west}%
	\inheritanchor[from=basic]{west north west}%
	\inheritanchor[from=basic]{east north east}%
	\inheritanchor[from=basic]{east south east}%
	\inheritanchor[from=basic]{north north west}%
	\inheritanchor[from=basic]{north north east}%
	\inheritanchor[from=basic]{south south west}%
	\inheritanchor[from=basic]{south south east}%
	
	\inheritanchorborder[from=basic]	
	\inheritbackgroundpath[from=basic]

    \pgfutil@g@addto@macro\pgf@sh@s@lensshape{%
        \pgfmathsetcount{\portcount}{0}
        \pgfmathloop%
        \ifnum\the\portcount<\nport
	        \pgfutil@ifundefined{pgf@anchor@lensshape@in\the\portcount}{
		        \expandafter\xdef\csname pgf@anchor@lensshape@in\the\portcount\endcsname{%
		            \noexpand\lensshape@port[\the\portcount]
		        }%
		    }{}%
	        \ifnum\the\portcount=0
    		    \pgfutil@ifundefined{pgf@anchor@lensshape@in}{%
		        \expandafter\xdef\csname pgf@anchor@lensshape@in\endcsname{%
		            \noexpand\lensshape@port[\the\portcount]
		        }%
		        }{}%
		    \fi
		    \pgfutil@ifundefined{pgf@anchor@lensshape@out\the\portcount}{%
		        \expandafter\xdef\csname pgf@anchor@lensshape@out\the\portcount\endcsname{%
		            \noexpand\lensshape@port[\the\portcount]
		        }%
		    }{}%
	        \ifnum\the\portcount=0
    		    \pgfutil@ifundefined{pgf@anchor@lensshape@out}{%
		        \expandafter\xdef\csname pgf@anchor@lensshape@out\endcsname{%
		            \noexpand\lensshape@port[\the\portcount]
		        }%
		        }{}%
		    \fi
	        \pgfmathaddtocount{\portcount}{1}	
	        \repeatpgfmathloop
	}
}

\def\lensshape@port[#1]{
    \northeast \pgf@xa=\pgf@x \pgf@ya=\pgf@y
    \southwest \pgf@xb=\pgf@x \pgf@yb=\pgf@y
    
	\pgfmathsetlength{\pgf@x}{0.5*\pgf@xa + 0.5*\pgf@xb}%
    \pgf@yc=\pgf@ya \advance\pgf@yc by -\pgf@yb	
    \pgfmathsetlength{\pgf@y}{\pgf@ya-(#1 + 0.5)*(\pgf@yc/\nport)}%
}

\pgfaddtoshape{lensshape}{
	\anchorlet{c}{center}		
	\anchorlet{n}{north}
	\anchorlet{e}{east}
	\anchorlet{s}{south}
	\anchorlet{w}{west}			
	\anchorlet{se}{south east}
	\anchorlet{sw}{south west}
	\anchorlet{ne}{north east}
	\anchorlet{nw}{north west}
	\anchorlet{wsw}{west south west}
	\anchorlet{wnw}{west north west}
	\anchorlet{ene}{east north east}
	\anchorlet{ese}{east south east}
	\anchorlet{nnw}{north north west}
	\anchorlet{nne}{north north east}
	\anchorlet{ssw}{south south west}
	\anchorlet{sse}{south south east}
}

\tikzset{
	/tikz/lenskeys/.cd,
	height/.initial=1,
	width/.initial=0.3,	
	color/.initial=O,
	rotation/.initial=0,
	linestyle/.initial={linestyle, inner sep=0.5mm},
	nport/.initial=1,
	/tikz/lens/.code={
		\pgfqkeys{/tikz/lenskeys}{#1}%
		\tikzset{/tikz/lenskeys/drawer/.expanded=%
			{\pgfkeysvalueof{/tikz/lenskeys/height}}%
			{\pgfkeysvalueof{/tikz/lenskeys/width}}%
			{\pgfkeysvalueof{/tikz/lenskeys/color}}%
			{\pgfkeysvalueof{/tikz/lenskeys/linestyle}}%
			{\pgfkeysvalueof{/tikz/lenskeys/rotation}}%
		}
	},
	/tikz/lenskeys/drawer/.code n args={5}{%
		\tikzset{
			lensshape,
			rotate=#5,
			minimum height=#1*\NODESIZE,
			minimum width=#2*\NODESIZE,
			append after command={
				\pgfextra{\let\bdr=\tikzlastnode%
					\draw[---, #3, #4, rounded corners = 0] (\bdr.n) to [in=120+#5, out=240+#5] (\bdr.s) to [in=-60+#5, out=60+#5] (\bdr.n) -- cycle;
				}
			}
		}
	},
}

\newcount\portcount

\pgfdeclareshape{mirrorshape}{
	\savedmacro\nport{
		\edef\nport{\pgfkeysvalueof{/tikz/mirrorkeys/nport}}%
	}
	\inheritsavedanchors[from=basic] 
	
	\inheritanchor[from=basic]{center}
	\inheritanchor[from=basic]{mid}		
	\inheritanchor[from=basic]{base}	
	\inheritanchor[from=basic]{north}
	\inheritanchor[from=basic]{south}		
	\inheritanchor[from=basic]{west}		
	\inheritanchor[from=basic]{mid west}				
	\inheritanchor[from=basic]{base west}		
	\inheritanchor[from=basic]{north west}		
	\inheritanchor[from=basic]{south west}		
	\inheritanchor[from=basic]{east}
	\inheritanchor[from=basic]{mid east}
	\inheritanchor[from=basic]{base east}	
	\inheritanchor[from=basic]{north east}
	\inheritanchor[from=basic]{south east}
	\inheritanchor[from=basic]{west south west}%
	\inheritanchor[from=basic]{west north west}%
	\inheritanchor[from=basic]{east north east}%
	\inheritanchor[from=basic]{east south east}%
	\inheritanchor[from=basic]{north north west}%
	\inheritanchor[from=basic]{north north east}%
	\inheritanchor[from=basic]{south south west}%
	\inheritanchor[from=basic]{south south east}%
	
	\inheritanchorborder[from=basic]	
	\inheritbackgroundpath[from=basic]

    \pgfutil@g@addto@macro\pgf@sh@s@mirrorshape{%
        \pgfmathsetcount{\portcount}{0}
        \pgfmathloop%
        \ifnum\the\portcount<\nport
	        \pgfutil@ifundefined{pgf@anchor@mirrorshape@in\the\portcount}{
		        \expandafter\xdef\csname pgf@anchor@mirrorshape@in\the\portcount\endcsname{%
		            \noexpand\mirrorshape@port[\the\portcount]
		        }%
		    }{}%
	        \ifnum\the\portcount=0
    		    \pgfutil@ifundefined{pgf@anchor@mirrorshape@in}{%
		        \expandafter\xdef\csname pgf@anchor@mirrorshape@in\endcsname{%
		            \noexpand\mirrorshape@port[\the\portcount]
		        }%
		        }{}%
		    \fi
		    \pgfutil@ifundefined{pgf@anchor@mirrorshape@out\the\portcount}{%
		        \expandafter\xdef\csname pgf@anchor@mirrorshape@out\the\portcount\endcsname{%
		            \noexpand\mirrorshape@port[\the\portcount]
		        }%
		    }{}%
	        \ifnum\the\portcount=0
    		    \pgfutil@ifundefined{pgf@anchor@mirrorshape@out}{%
		        \expandafter\xdef\csname pgf@anchor@mirrorshape@out\endcsname{%
		            \noexpand\mirrorshape@port[\the\portcount]
		        }%
		        }{}%
		    \fi
	        \pgfmathaddtocount{\portcount}{1}	
	        \repeatpgfmathloop
	}
}

\def\mirrorshape@port[#1]{
    \northeast \pgf@xa=\pgf@x \pgf@ya=\pgf@y
    \southwest \pgf@xb=\pgf@x \pgf@yb=\pgf@y
    
	\pgf@x=\pgf@xb
    \pgf@yc=\pgf@ya \advance\pgf@yc by -\pgf@yb	
    \pgfmathsetlength{\pgf@y}{\pgf@ya-(#1 + 0.5)*(\pgf@yc/\nport)}%
}

\pgfaddtoshape{mirrorshape}{
	\anchorlet{c}{center}		
	\anchorlet{n}{north}
	\anchorlet{e}{east}
	\anchorlet{s}{south}
	\anchorlet{w}{west}			
	\anchorlet{se}{south east}
	\anchorlet{sw}{south west}
	\anchorlet{ne}{north east}
	\anchorlet{nw}{north west}
	\anchorlet{wsw}{west south west}
	\anchorlet{wnw}{west north west}
	\anchorlet{ene}{east north east}
	\anchorlet{ese}{east south east}
	\anchorlet{nnw}{north north west}
	\anchorlet{nne}{north north east}
	\anchorlet{ssw}{south south west}
	\anchorlet{sse}{south south east}
}

\tikzset{
	/tikz/mirrorkeys/.cd,
	height/.initial=1,
	width/.initial=0.15,	
	color/.initial=O,
	rotation/.initial=0,
	linestyle/.initial={linestyle, inner sep=0.5mm},
	nport/.initial=1,
	nlines/.initial=5,
	/tikz/mirror/.code={
		\pgfqkeys{/tikz/mirrorkeys}{#1}%
		\tikzset{/tikz/mirrorkeys/drawer/.expanded=%
			{\pgfkeysvalueof{/tikz/mirrorkeys/height}}%
			{\pgfkeysvalueof{/tikz/mirrorkeys/width}}%
			{\pgfkeysvalueof{/tikz/mirrorkeys/color}}%
			{\pgfkeysvalueof{/tikz/mirrorkeys/linestyle}}%
			{\pgfkeysvalueof{/tikz/mirrorkeys/rotation}}%
			{\pgfkeysvalueof{/tikz/mirrorkeys/nlines}}%
		}
	},
	/tikz/mirrorkeys/drawer/.code n args={6}{%
		\tikzset{
			mirrorshape,
			rotate=#5,
			minimum height=#1*\NODESIZE,
			minimum width=#2*\NODESIZE,
			append after command={
				\pgfextra{\let\bdr=\tikzlastnode%
					\draw[---, #3, #4] (\bdr.nw) to (\bdr.sw){};
					\foreach \nline [evaluate=\nline as \linepos using (\nline-0.8)/(#6-0.6)] in {1,...,#6}{
						\draw[---, #3, #4] ($(\bdr.nw)!\linepos!(\bdr.sw)$) to +(#5-45:#2*1.41421356237*\NODESIZE pt){};
					}
				}
			}
		}
	},
}
\pgfdeclareshape{muxshape}{
	\savedmacro\nports{
		\edef\nports{\pgfkeysvalueof{/tikz/muxkeys/nports}}%
	}
	\savedmacro\direction{
		\edef\direction{\pgfkeysvalueof{/tikz/muxkeys/direction}}%
	}
	\savedmacro\inverted{
		\edef\inverted{\pgfkeysvalueof{/tikz/muxkeys/inverted}}%
	}
	\savedmacro\ninports{
		\ifnum\inverted=0
			\edef\ninports{\nports}
		\else
			\edef\ninports{1}%
		\fi
	}
	\savedmacro\noutports{
		\ifnum\inverted=0
			\edef\noutports{1}%
		\else
			\edef\noutports{\nports}%
		\fi
	}
	\inheritsavedanchors[from=basic]

	\inheritanchor[from=basic]{center}
	\inheritanchor[from=basic]{mid}
	\inheritanchor[from=basic]{base}
	\inheritanchor[from=basic]{north}
	\inheritanchor[from=basic]{south}
	\inheritanchor[from=basic]{west}
	\inheritanchor[from=basic]{mid west}
	\inheritanchor[from=basic]{base west}
	\inheritanchor[from=basic]{north west}
	\inheritanchor[from=basic]{south west}
	\inheritanchor[from=basic]{east}
	\inheritanchor[from=basic]{mid east}
	\inheritanchor[from=basic]{base east}
	\inheritanchor[from=basic]{north east}
	\inheritanchor[from=basic]{south east}
	\inheritanchor[from=basic]{west south west}%
	\inheritanchor[from=basic]{west north west}%
	\inheritanchor[from=basic]{east north east}%
	\inheritanchor[from=basic]{east south east}%
	\inheritanchor[from=basic]{north north west}%
	\inheritanchor[from=basic]{north north east}%
	\inheritanchor[from=basic]{south south west}%
	\inheritanchor[from=basic]{south south east}%

	\inheritanchorborder[from=basic]
	\inheritbackgroundpath[from=basic]

	\pgfutil@g@addto@macro\pgf@sh@s@muxshape{%
		\pgfmathsetcount{\portcount}{0}
		\pgfmathloop%
		\ifnum\the\portcount<\nports
		\ifnum\the\portcount<\ninports
			\pgfutil@ifundefined{pgf@anchor@muxshape@in\the\portcount}{
				\expandafter\xdef\csname pgf@anchor@muxshape@in\the\portcount\endcsname{%
					\noexpand\muxshape@port[\the\portcount]{0}
				}%
			}{}%
			\ifnum\the\portcount=0
				\pgfutil@ifundefined{pgf@anchor@muxshape@in}{%
					\expandafter\xdef\csname pgf@anchor@muxshape@in\endcsname{%
						\noexpand\muxshape@port[\the\portcount]{0}
					}%
				}{}%
			\fi
		\fi
		\ifnum\the\portcount<\noutports
			\pgfutil@ifundefined{pgf@anchor@muxshape@out\the\portcount}{%
				\expandafter\xdef\csname pgf@anchor@muxshape@out\the\portcount\endcsname{%
					\noexpand\muxshape@port[\the\portcount]{1}
				}%
			}{}%
			\ifnum\the\portcount=0
				\pgfutil@ifundefined{pgf@anchor@muxshape@out}{%
					\expandafter\xdef\csname pgf@anchor@muxshape@out\endcsname{%
						\noexpand\muxshape@port[\the\portcount]{1}
					}%
				}{}%
			\fi
		\fi
		\pgfmathaddtocount{\portcount}{1}	
		\repeatpgfmathloop
	}
}

\def\muxshape@port[#1]#2{
	\northeast \pgf@xa=\pgf@x \pgf@ya=\pgf@y
	\southwest \pgf@xb=\pgf@x \pgf@yb=\pgf@y

	\ifnum#2=0
		\ifnum\inverted=0
			\def\chooseports{0}
		\else
			\def\chooseports{1}
		\fi
	\else
		\ifnum\inverted=0
			\def\chooseports{1}
		\else
			\def\chooseports{0}
		\fi
	\fi

	\ifnum\chooseports=0	
		\if\direction\direce
			\pgf@x=\pgf@xb
			\pgf@yc=\pgf@ya \advance\pgf@yc by -\pgf@yb	
			\pgfmathsetlength{\pgf@y}{\pgf@ya-(#1 + 0.5)*(\pgf@yc/\nports)}%
		\fi
		\if\direction\direcw
			\pgf@x=\pgf@xa
			\pgf@yc=\pgf@ya \advance\pgf@yc by -\pgf@yb	
			\pgfmathsetlength{\pgf@y}{\pgf@ya-(#1 + 0.5)*(\pgf@yc/\nports)}%
		\fi
		\if\direction\direcn
			\pgf@y=\pgf@yb
			\pgf@xc=\pgf@xa \advance\pgf@xc by -\pgf@xb	
			\pgfmathsetlength{\pgf@x}{\pgf@xb+(#1 + 0.5)*(\pgf@xc/\nports)}%
		\fi
		\if\direction\direcs
			\pgf@y=\pgf@ya
			\pgf@xc=\pgf@xa \advance\pgf@xc by -\pgf@xb	
			\pgfmathsetlength{\pgf@x}{\pgf@xb+(#1 + 0.5)*(\pgf@xc/\nports)}%
		\fi
	\else	
		\if\direction\direce
			\pgf@x=\pgf@xa
			\pgf@yc=\pgf@ya \advance\pgf@yc by -\pgf@yb	
			\pgfmathsetlength{\pgf@y}{\pgf@ya-0.5\pgf@yc}%
		\fi
		\if\direction\direcw
			\pgf@x=\pgf@xb
			\pgf@yc=\pgf@ya \advance\pgf@yc by -\pgf@yb	
			\pgfmathsetlength{\pgf@y}{\pgf@ya-0.5\pgf@yc}%
		\fi
		\if\direction\direcn
			\pgf@y=\pgf@ya
			\pgf@xc=\pgf@xa \advance\pgf@xc by -\pgf@xb	
			\pgfmathsetlength{\pgf@x}{\pgf@xa-0.5\pgf@xc}%
		\fi
		\if\direction\direcs
			\pgf@y=\pgf@yb
			\pgf@xc=\pgf@xa \advance\pgf@xc by -\pgf@xb	
			\pgfmathsetlength{\pgf@x}{\pgf@xa-0.5\pgf@xc}%
		\fi
	\fi
}

\pgfaddtoshape{muxshape}{
	\anchorlet{c}{center}
	\anchorlet{n}{north}
	\anchorlet{e}{east}
	\anchorlet{s}{south}
	\anchorlet{w}{west}
	\anchorlet{se}{south east}
	\anchorlet{sw}{south west}
	\anchorlet{ne}{north east}
	\anchorlet{nw}{north west}
	\anchorlet{wsw}{west south west}
	\anchorlet{wnw}{west north west}
	\anchorlet{ene}{east north east}
	\anchorlet{ese}{east south east}
	\anchorlet{nnw}{north north west}
	\anchorlet{nne}{north north east}
	\anchorlet{ssw}{south south west}
	\anchorlet{sse}{south south east}
}

\tikzset{
/tikz/muxkeys/.cd,
height/.initial=1,
width/.initial=0.5,
color/.initial=O,
direction/.initial=e,
linestyle/.initial={linestyle, rounded corners = 0},
nports/.initial=3,
inverted/.initial=0,	
smux/.initial=0,
angle/.initial=60,
fillgradient/.initial=O,
/tikz/mux/.code={
\pgfqkeys{/tikz/muxkeys}{#1}%
\tikzset{/tikz/muxkeys/drawer/.expanded=%
\if\pgfkeysvalueof{/tikz/muxkeys/direction}e
	{\pgfkeysvalueof{/tikz/muxkeys/width}}%
	{\pgfkeysvalueof{/tikz/muxkeys/height}}%
	{0}%
	{-90}%
\fi
\if\pgfkeysvalueof{/tikz/muxkeys/direction}w
	{\pgfkeysvalueof{/tikz/muxkeys/width}}%
	{\pgfkeysvalueof{/tikz/muxkeys/height}}%
	{0}%
	{90}%
\fi
\if\pgfkeysvalueof{/tikz/muxkeys/direction}n
	{\pgfkeysvalueof{/tikz/muxkeys/width}}%
	{\pgfkeysvalueof{/tikz/muxkeys/height}}%
	{1}%
	{0}%
\fi
\if\pgfkeysvalueof{/tikz/muxkeys/direction}s
	{\pgfkeysvalueof{/tikz/muxkeys/width}}%
	{\pgfkeysvalueof{/tikz/muxkeys/height}}%
	{1}%
	{180}%
\fi
{\pgfkeysvalueof{/tikz/muxkeys/color}}%
{\pgfkeysvalueof{/tikz/muxkeys/linestyle}}%
{\pgfkeysvalueof{/tikz/muxkeys/smux}}%
{\pgfkeysvalueof{/tikz/muxkeys/angle}}%
{\pgfkeysvalueof{/tikz/muxkeys/fillgradient}}%
}
},
/tikz/muxkeys/drawer/.code n args={9}{%
		\tikzset{
			muxshape,
			#6,
			#5,
			minimum height=
			\ifnum#3>0	
				#1*\NODESIZE
			\else
				#2*\NODESIZE
			\fi
			,minimum width=
			\ifnum#3>0
				#2*\NODESIZE
			\else
				#1*\NODESIZE
			\fi
			,append after command={
					\pgfextra{\let\bdr=\tikzlastnode%
						\node[trapezium, line width = \NODETHICKNESS, minimum height=#1*\NODESIZE, minimum width=#2*\NODESIZE, trapezium stretches=true, rotate=#4, trapezium angle=#8, inner sep=0.001mm] at (\bdr) (trap) {};	
						\ifnum#7=0
							\draw[#9, #5, #6] (trap.bottom left corner) to (trap.top left corner) to (trap.top right corner) to (trap.bottom right corner) to cycle;
						\else
							\draw[#9, #5, #6] (trap.bottom left corner) to[in=#4-90, out=#4+90] (trap.top left corner) to (trap.top right corner) to[in=#4+90,out=#4-90] (trap.bottom right corner) to cycle;
						\fi
					}
				}
		}
	},
}
\newcount\portcount

\pgfdeclareshape{polswitchshape}{
	\savedmacro\nin{
		\edef\nin{\pgfkeysvalueof{/tikz/polswitchkeys/nin}}%
	}
	\savedmacro\nout{
		\edef\nout{\pgfkeysvalueof{/tikz/polswitchkeys/nout}}%
	}
	\savedmacro\direction{
		\edef\direction{\pgfkeysvalueof{/tikz/polswitchkeys/direction}}%
	}
	\inheritsavedanchors[from=basic] 
	
	\inheritanchor[from=basic]{center}
	\inheritanchor[from=basic]{mid}		
	\inheritanchor[from=basic]{base}	
	\inheritanchor[from=basic]{north}
	\inheritanchor[from=basic]{south}		
	\inheritanchor[from=basic]{west}		
	\inheritanchor[from=basic]{mid west}				
	\inheritanchor[from=basic]{base west}		
	\inheritanchor[from=basic]{north west}		
	\inheritanchor[from=basic]{south west}		
	\inheritanchor[from=basic]{east}
	\inheritanchor[from=basic]{mid east}
	\inheritanchor[from=basic]{base east}	
	\inheritanchor[from=basic]{north east}
	\inheritanchor[from=basic]{south east}
	\inheritanchor[from=basic]{west south west}%
	\inheritanchor[from=basic]{west north west}%
	\inheritanchor[from=basic]{east north east}%
	\inheritanchor[from=basic]{east south east}%
	\inheritanchor[from=basic]{north north west}%
	\inheritanchor[from=basic]{north north east}%
	\inheritanchor[from=basic]{south south west}%
	\inheritanchor[from=basic]{south south east}%
	
	\inheritanchorborder[from=basic]	
	\inheritbackgroundpath[from=basic]

    \pgfutil@g@addto@macro\pgf@sh@s@polswitchshape{%
        \pgfmathsetcount{\portcount}{0}
        \pgfmathloop%
        \ifnum\the\portcount<\nin
	        \pgfutil@ifundefined{pgf@anchor@polswitchshape@in\the\portcount}{
		        \expandafter\xdef\csname pgf@anchor@polswitchshape@in\the\portcount\endcsname{%
		            \noexpand\polswitchshape@port[\the\portcount]{0}
		        }%
		    }{}%
	        \ifnum\the\portcount=0
    		    \pgfutil@ifundefined{pgf@anchor@polswitchshape@in}{%
		        \expandafter\xdef\csname pgf@anchor@polswitchshape@in\endcsname{%
		            \noexpand\polswitchshape@port[\the\portcount]{0}
		        }%
		        }{}%
		    \fi
	        \pgfmathaddtocount{\portcount}{1}	
	        \repeatpgfmathloop
        \pgfmathsetcount{\portcount}{0}
        \pgfmathloop%
    	\ifnum\the\portcount<\nout
	        \pgfutil@ifundefined{pgf@anchor@polswitchshape@out\the\portcount}{%
		        \expandafter\xdef\csname pgf@anchor@polswitchshape@out\the\portcount\endcsname{%
		            \noexpand\polswitchshape@port[\the\portcount]{1}
		        }%
		    }{}%
	        \ifnum\the\portcount=0
    		    \pgfutil@ifundefined{pgf@anchor@polswitchshape@out}{%
		        \expandafter\xdef\csname pgf@anchor@polswitchshape@out\endcsname{%
		            \noexpand\polswitchshape@port[\the\portcount]{1}
		        }%
		        }{}%
		    \fi
	        \pgfmathaddtocount{\portcount}{1}	
	        \repeatpgfmathloop
	}
}

\def\polswitchshape@port[#1]#2{
    \northeast \pgf@xa=\pgf@x \pgf@ya=\pgf@y
    \southwest \pgf@xb=\pgf@x \pgf@yb=\pgf@y
    
    \ifnum#2=0	
	    \if\direction\direce	
	    	\pgf@x=\pgf@xb
		    \pgf@yc=\pgf@ya \advance\pgf@yc by -\pgf@yb	
		    \pgfmathsetlength{\pgf@y}{\pgf@ya-(#1 + 0.5)*(\pgf@yc/\nin)}%
	    \fi
	    \if\direction\direcw
	    	\pgf@x=\pgf@xa
		    \pgf@yc=\pgf@ya \advance\pgf@yc by -\pgf@yb	
		    \pgfmathsetlength{\pgf@y}{\pgf@ya-(#1 + 0.5)*(\pgf@yc/\nin)}%
	    \fi
	    \if\direction\direcn
	    	\pgf@y=\pgf@yb
		    \pgf@xc=\pgf@xa \advance\pgf@xc by -\pgf@xb	
		    \pgfmathsetlength{\pgf@x}{\pgf@xb+(#1 + 0.5)*(\pgf@xc/\nin)}%
	    \fi
	    \if\direction\direcs
	    	\pgf@y=\pgf@ya
		    \pgf@xc=\pgf@xa \advance\pgf@xc by -\pgf@xb	
		    \pgfmathsetlength{\pgf@x}{\pgf@xb+(#1 + 0.5)*(\pgf@xc/\nin)}%
	    \fi
	\else	
	    \if\direction\direce	
	    	\pgf@x=\pgf@xa
		    \pgf@yc=\pgf@ya \advance\pgf@yc by -\pgf@yb	
		    \pgfmathsetlength{\pgf@y}{\pgf@ya-(#1 + 0.5)*(\pgf@yc/\nout)}%
	    \fi
	    \if\direction\direcw
	    	\pgf@x=\pgf@xb
		    \pgf@yc=\pgf@ya \advance\pgf@yc by -\pgf@yb	
		    \pgfmathsetlength{\pgf@y}{\pgf@ya-(#1 + 0.5)*(\pgf@yc/\nout)}%
	    \fi
	    \if\direction\direcn
	    	\pgf@y=\pgf@ya
		    \pgf@xc=\pgf@xa \advance\pgf@xc by -\pgf@xb	
		    \pgfmathsetlength{\pgf@x}{\pgf@xb+(#1 + 0.5)*(\pgf@xc/\nout)}%
	    \fi
	    \if\direction\direcs
	    	\pgf@y=\pgf@yb
		    \pgf@xc=\pgf@xa \advance\pgf@xc by -\pgf@xb	
		    \pgfmathsetlength{\pgf@x}{\pgf@xb+(#1 + 0.5)*(\pgf@xc/\nout)}%
	    \fi
	\fi
}

\pgfaddtoshape{polswitchshape}{
	\anchorlet{c}{center}		
	\anchorlet{n}{north}
	\anchorlet{e}{east}
	\anchorlet{s}{south}
	\anchorlet{w}{west}			
	\anchorlet{se}{south east}
	\anchorlet{sw}{south west}
	\anchorlet{ne}{north east}
	\anchorlet{nw}{north west}
	\anchorlet{wsw}{west south west}
	\anchorlet{wnw}{west north west}
	\anchorlet{ene}{east north east}
	\anchorlet{ese}{east south east}
	\anchorlet{nnw}{north north west}
	\anchorlet{nne}{north north east}
	\anchorlet{ssw}{south south west}
	\anchorlet{sse}{south south east}
}

\tikzset{
	/tikz/polswitchkeys/.cd,
	size/.initial=1,
	color/.initial=O,
	direction/.initial=e,
	linestyle/.initial={linestyle, inner sep=0.5mm},
	nin/.initial=1,	
	nout/.initial=1, 
	/tikz/polswitch/.code={
		\pgfqkeys{/tikz/polswitchkeys}{#1}%
		\tikzset{/tikz/polswitchkeys/drawer/.expanded=%
			{\pgfkeysvalueof{/tikz/polswitchkeys/size}}%
			{\pgfkeysvalueof{/tikz/polswitchkeys/color}}%
			{\pgfkeysvalueof{/tikz/polswitchkeys/linestyle}}%
			{\pgfkeysvalueof{/tikz/polswitchkeys/nout}}%
			\if\pgfkeysvalueof{/tikz/polswitchkeys/direction}e
				{0}%
			\fi
			\if\pgfkeysvalueof{/tikz/polswitchkeys/direction}w
				{0}%
			\fi
			\if\pgfkeysvalueof{/tikz/polswitchkeys/direction}n
				{1}%
			\fi
			\if\pgfkeysvalueof{/tikz/polswitchkeys/direction}s
				{1}%
			\fi
			{\pgfkeysvalueof{/tikz/polswitchkeys/direction}}%
		}
	},
	/tikz/polswitchkeys/drawer/.code n args={6}{%
		\tikzset{
			polswitchshape,
			draw,
			minimum height = #1*\NODESIZE,
			minimum width = #1*\NODESIZE,
			#2,
			#3,
			append after command={
				\pgfextra{\let\bdr=\tikzlastnode%
						\node[coordinate] at ($(\bdr.in)!0.25!(\bdr.out)$) (circlein){};
						\node[coordinate] at ($(\bdr.in)!0.75!(\bdr.out)$) (circleout){};

						\draw[---, #2, #3, fill] (\bdr.in) to (circlein) circle (0.05);
						\draw[---, #2, #3, fill] (\bdr.out) to (circleout) circle (0.05);

						\node[coordinate] at ($(\bdr.in)!0.6!(\bdr.out)$) (circlemiddle){};
						\ifnum#5>0
							\node[coordinate] at ($(circlemiddle)!0.5!(circlemiddle -| \bdr.e)$) (circletopcor){};
							\node[coordinate] at ($(circlemiddle)!0.5!(circlemiddle -| \bdr.w)$) (circlebotcor){};
						\else
							\node[coordinate] at ($(circlemiddle)!0.5!(circlemiddle |- \bdr.n)$) (circletopcor){};
							\node[coordinate] at ($(circlemiddle)!0.5!(circlemiddle |- \bdr.s)$) (circlebotcor){};
						\fi

						\node[draw, circle, #2, #3, minimum size=0.3*\FNODESIZE] at (circletopcor) (circletop){};
						\node[draw, circle, #2, #3, minimum size=0.3*\FNODESIZE] at (circlebotcor) (circlebot){};
						\draw[-->, #2, #3] (circletop.south) to (circletop.north){};
						\draw[-->, #2, #3] (circlebot.west) to (circlebot.east){};

						\if#6e
							\draw[-->, #2, #3] ([xshift=-0.05*\FNODESIZE]circletop.south west) to [out=-120, in=120] ([xshift=-0.05*\FNODESIZE]circlebot.north west){};
						\fi
						\if#6w
							\draw[-->, #2, #3] ([xshift=0.05*\FNODESIZE]circletop.south east) to [out=-60, in=60] ([xshift=0.05*\FNODESIZE]circlebot.north east){};
						\fi
						\if#6s
							\draw[-->, #2, #3] ([yshift=0.05*\FNODESIZE]circletop.north west) to [out=150, in=30] ([yshift=0.05*\FNODESIZE]circlebot.north east){};
						\fi
						\if#6n
							\draw[-->, #2, #3] ([yshift=-0.05*\FNODESIZE]circletop.south west) to [out=-150, in=-30] ([yshift=-0.05*\FNODESIZE]circlebot.south east){};
						\fi

				}
			}
		}
	},
}
\pgfdeclareshape{pdshape}{
	\savedmacro\direction{
		\edef\direction{\pgfkeysvalueof{/tikz/pdkeys/direction}}%
	}
	\saveddimen\minwidth{
		\pgfmathsetlength\pgf@x{\pgfshapeminwidth}%
	}
	\saveddimen\minheight{
		\pgfmathsetlength\pgf@x{\pgfshapeminheight}%
	}
	\inheritsavedanchors[from=basic]

	\inheritanchor[from=basic]{center}
	\inheritanchor[from=basic]{mid}
	\inheritanchor[from=basic]{base}
	\inheritanchor[from=basic]{north}
	\inheritanchor[from=basic]{south}
	\inheritanchor[from=basic]{west}
	\inheritanchor[from=basic]{mid west}
	\inheritanchor[from=basic]{base west}
	\inheritanchor[from=basic]{north west}
	\inheritanchor[from=basic]{south west}
	\inheritanchor[from=basic]{east}
	\inheritanchor[from=basic]{mid east}
	\inheritanchor[from=basic]{base east}
	\inheritanchor[from=basic]{north east}
	\inheritanchor[from=basic]{south east}
	\inheritanchor[from=basic]{west south west}%
	\inheritanchor[from=basic]{west north west}%
	\inheritanchor[from=basic]{east north east}%
	\inheritanchor[from=basic]{east south east}%
	\inheritanchor[from=basic]{north north west}%
	\inheritanchor[from=basic]{north north east}%
	\inheritanchor[from=basic]{south south west}%
	\inheritanchor[from=basic]{south south east}%

	\inheritanchorborder[from=basic]
	\inheritbackgroundpath[from=basic]

	\pgfutil@g@addto@macro\pgf@sh@s@pdshape{%
		\pgfutil@ifundefined{pgf@anchor@pdshape@in0}{
			\expandafter\xdef\csname pgf@anchor@pdshape@in0\endcsname{%
				\noexpand\pdshape@port{0}
			}%
		}{}%
		\pgfutil@ifundefined{pgf@anchor@pdshape@in}{
			\expandafter\xdef\csname pgf@anchor@pdshape@in\endcsname{%
				\noexpand\pdshape@port{0}
			}%
		}{}%
		\pgfutil@ifundefined{pgf@anchor@pdshape@out0}{
			\expandafter\xdef\csname pgf@anchor@pdshape@out0\endcsname{%
				\noexpand\pdshape@port{1}
			}%
		}{}%
		\pgfutil@ifundefined{pgf@anchor@pdshape@out}{
			\expandafter\xdef\csname pgf@anchor@pdshape@out\endcsname{%
				\noexpand\pdshape@port{1}
			}%
		}{}%
	}
}

\def\pdshape@port#1{
	\northeast	

	\ifnum#1=0	
		\if\direction\direce
			\pgf@x=-\pgf@x
			\pgf@ya= \pgf@y
			\pgfmathsetlength{\pgf@y}{\pgf@ya-0.5*\minheight}%
		\fi
		\if\direction\direcw
			\pgf@x=\pgf@x
			\pgf@ya= \pgf@y
			\pgfmathsetlength{\pgf@y}{\pgf@ya-0.5*\minheight}%
		\fi
		\if\direction\direcn
			\pgf@y=-\pgf@y
			\pgf@xa=\pgf@x
			\pgfmathsetlength{\pgf@x}{\pgf@xa-0.5*\minwidth}%
		\fi
		\if\direction\direcs
			\pgf@y=\pgf@y
			\pgf@xa= \pgf@x
			\pgfmathsetlength{\pgf@x}{\pgf@xa-0.5*\minwidth}%
		\fi
	\else	
		\if\direction\direce
			\pgf@x=\pgf@x
			\pgf@ya= \pgf@y
			\pgfmathsetlength{\pgf@y}{\pgf@ya-0.5*\minheight}%
		\fi
		\if\direction\direcw
			\pgf@x=-\pgf@x
			\pgf@ya= \pgf@y
			\pgfmathsetlength{\pgf@y}{\pgf@ya-0.5*\minheight}%
		\fi
		\if\direction\direcn
			\pgf@y=\pgf@y
			\pgf@xa= \pgf@x
			\pgfmathsetlength{\pgf@x}{\pgf@xa-0.5*\minwidth}%
		\fi
		\if\direction\direcs
			\pgf@y=-\pgf@y
			\pgf@xa= \pgf@x
			\pgfmathsetlength{\pgf@x}{\pgf@xa-0.5*\minwidth}%
		\fi
	\fi
}

\pgfaddtoshape{pdshape}{
	\anchorlet{c}{center}
	\anchorlet{n}{north}
	\anchorlet{e}{east}
	\anchorlet{s}{south}
	\anchorlet{w}{west}
	\anchorlet{se}{south east}
	\anchorlet{sw}{south west}
	\anchorlet{ne}{north east}
	\anchorlet{nw}{north west}
	\anchorlet{wsw}{west south west}
	\anchorlet{wnw}{west north west}
	\anchorlet{ene}{east north east}
	\anchorlet{ese}{east south east}
	\anchorlet{nnw}{north north west}
	\anchorlet{nne}{north north east}
	\anchorlet{ssw}{south south west}
	\anchorlet{sse}{south south east}
}

\tikzset{
/tikz/pdkeys/.cd,
size/.initial=0.5,
color/.initial=EO,
direction/.initial=e,
linestyle/.initial={linestyle},
fillgradient/.initial=O,
/tikz/pd/.code={
		\pgfqkeys{/tikz/pdkeys}{#1}%
		\tikzset{/tikz/pdkeys/drawer/.expanded=%
				{\pgfkeysvalueof{/tikz/pdkeys/direction}}%
				{\pgfkeysvalueof{/tikz/pdkeys/size}}%
				{\pgfkeysvalueof{/tikz/pdkeys/color}}%
				{\pgfkeysvalueof{/tikz/pdkeys/linestyle}}%
				{\pgfkeysvalueof{/tikz/pdkeys/fillgradient}}%
		}
	},
/tikz/pdkeys/drawer/.code n args={5}{%
\tikzset{
pdshape,
minimum height=#2*\NODESIZE,
minimum width=#2*\NODESIZE,
#3,
#4,
draw,
append after command={
\pgfextra{\let\bdr=\tikzlastnode%
\node[#5, fit=(\bdr.nw)(\bdr.se)] (boxgradient){};
\draw[---,#3] ($(\bdr.s)!.1!(\bdr.n)$) to ($(\bdr.s)!.9!(\bdr.n)$);
\fill[#3] ({$(\bdr.s)!.3!(\bdr.n)$} -| {$(\bdr.w)!.3!(\bdr.e)$}) to ($(\bdr.s)!.7!(\bdr.n)$) to ({$(\bdr.s)!.3!(\bdr.n)$} -| {$(\bdr.w)!.7!(\bdr.e)$}) to cycle;
\draw[---,#3] ({$(\bdr.s)!.7!(\bdr.n)$} -| {$(\bdr.w)!.35!(\bdr.e)$}) to ({$(\bdr.s)!.7!(\bdr.n)$} -| {$(\bdr.w)!.65!(\bdr.e)$});
}
}
}
},
}
\pgfdeclareshape{pbsshape}{
	\savedmacro\direction{
		\edef\direction{\pgfkeysvalueof{/tikz/pbskeys/direction}}%
	}
	\saveddimen\minwidth{
		\pgfmathsetlength\pgf@x{\pgfshapeminwidth}%
	}
	\saveddimen\minheight{
		\pgfmathsetlength\pgf@x{\pgfshapeminheight}%
	}
	\savedmacro\nport{
		\edef\nport{\pgfkeysvalueof{/tikz/pbskeys/nport}}%
	}
	\inheritsavedanchors[from=basic]

	\inheritanchor[from=basic]{center}
	\inheritanchor[from=basic]{mid}
	\inheritanchor[from=basic]{base}
	\inheritanchor[from=basic]{north}
	\inheritanchor[from=basic]{south}
	\inheritanchor[from=basic]{west}
	\inheritanchor[from=basic]{mid west}
	\inheritanchor[from=basic]{base west}
	\inheritanchor[from=basic]{north west}
	\inheritanchor[from=basic]{south west}
	\inheritanchor[from=basic]{east}
	\inheritanchor[from=basic]{mid east}
	\inheritanchor[from=basic]{base east}
	\inheritanchor[from=basic]{north east}
	\inheritanchor[from=basic]{south east}
	\inheritanchor[from=basic]{west south west}%
	\inheritanchor[from=basic]{west north west}%
	\inheritanchor[from=basic]{east north east}%
	\inheritanchor[from=basic]{east south east}%
	\inheritanchor[from=basic]{north north west}%
	\inheritanchor[from=basic]{north north east}%
	\inheritanchor[from=basic]{south south west}%
	\inheritanchor[from=basic]{south south east}%

	\inheritanchorborder[from=basic]
	\inheritbackgroundpath[from=basic]

	\pgfutil@g@addto@macro\pgf@sh@s@pbsshape{%
		\pgfutil@ifundefined{pgf@anchor@pbsshape@in0}{
			\expandafter\xdef\csname pgf@anchor@pbsshape@in0\endcsname{%
				\noexpand\pbsshape@port[0]{0}
			}%
		}{}%
		\pgfutil@ifundefined{pgf@anchor@pbsshape@in1}{
			\expandafter\xdef\csname pgf@anchor@pbsshape@in1\endcsname{%
				\noexpand\pbsshape@port[1]{0}
			}%
		}{}%
		\pgfutil@ifundefined{pgf@anchor@pbsshape@in}{
			\expandafter\xdef\csname pgf@anchor@pbsshape@in\endcsname{%
				\noexpand\pbsshape@port[0]{0}
			}%
		}{}%
		\pgfutil@ifundefined{pgf@anchor@pbsshape@out0}{
			\expandafter\xdef\csname pgf@anchor@pbsshape@out0\endcsname{%
				\noexpand\pbsshape@port[0]{1}
			}%
		}{}%
		\pgfutil@ifundefined{pgf@anchor@pbsshape@out1}{
			\expandafter\xdef\csname pgf@anchor@pbsshape@out1\endcsname{%
				\noexpand\pbsshape@port[1]{1}
			}%
		}{}%
		\pgfutil@ifundefined{pgf@anchor@pbsshape@out}{
			\expandafter\xdef\csname pgf@anchor@pbsshape@out\endcsname{%
				\noexpand\pbsshape@port[0]{1}
			}%
		}{}%
		\pgfmathsetcount{\portcount}{0}
		\pgfmathloop%
		\ifnum\the\portcount<\nport
		\pgfutil@ifundefined{pgf@anchor@pbsshape@p\the\portcount}{
			\expandafter\xdef\csname pgf@anchor@pbsshape@p\the\portcount\endcsname{%
				\noexpand\pbsshape@port[\the\portcount]{2}
			}%
		}{}%
		\pgfmathaddtocount{\portcount}{1}	
		\repeatpgfmathloop
	}
}

\def\pbsshape@port[#1]#2{
	\northeast	

	\ifnum#2=0	
		\if\direction\direce
			\ifnum#1=0
				\pgf@x=-\pgf@x
				\pgf@ya= \pgf@y
				\pgfmathsetlength{\pgf@y}{\pgf@ya-0.5*\minheight}%
			\else
				\pgf@x=0\pgf@x
				\pgf@ya= \pgf@y
				\pgfmathsetlength{\pgf@y}{\pgf@ya-\minheight}%
			\fi
		\fi
		\if\direction\direcw
			\ifnum#1=0
				\pgf@x=\pgf@x
				\pgf@ya= \pgf@y
				\pgfmathsetlength{\pgf@y}{\pgf@ya-0.5*\minheight}%
			\else
				\pgf@x=0\pgf@x
				\pgf@ya= \pgf@y
				\pgfmathsetlength{\pgf@y}{\pgf@ya}%
			\fi
		\fi
		\if\direction\direcn
			\ifnum#1=0
				\pgf@y=-\pgf@y
				\pgf@xa=\pgf@x
				\pgfmathsetlength{\pgf@x}{\pgf@xa-0.5*\minwidth}%
			\else
				\pgf@y=0\pgf@y
				\pgf@xa=\pgf@x
				\pgfmathsetlength{\pgf@x}{\pgf@xa-0*\minwidth}%
			\fi
		\fi
		\if\direction\direcs
			\ifnum#1=0
				\pgf@y=\pgf@y
				\pgf@xa= \pgf@x
				\pgfmathsetlength{\pgf@x}{\pgf@xa-0.5*\minwidth}%
			\else
				\pgf@y=0\pgf@y
				\pgf@xa= \pgf@x
				\pgfmathsetlength{\pgf@x}{\pgf@xa-1*\minwidth}%
			\fi
		\fi
	\else	
		\if\direction\direce
			\ifnum#1=0
				\pgf@x=\pgf@x
				\pgf@ya= \pgf@y
				\pgfmathsetlength{\pgf@y}{\pgf@ya-0.5*\minheight}%
			\else
				\pgf@x=0\pgf@x
				\pgf@ya= \pgf@y
				\pgfmathsetlength{\pgf@y}{\pgf@ya}%
			\fi
		\fi
		\if\direction\direcw
			\ifnum#1=0
				\pgf@x=-\pgf@x
				\pgf@ya= \pgf@y
				\pgfmathsetlength{\pgf@y}{\pgf@ya-0.5*\minheight}%
			\else
				\pgf@x=0\pgf@x
				\pgf@ya= \pgf@y
				\pgfmathsetlength{\pgf@y}{\pgf@ya-\minheight}%
			\fi
		\fi
		\if\direction\direcn
			\ifnum#1=0
				\pgf@y=\pgf@y
				\pgf@xa= \pgf@x
				\pgfmathsetlength{\pgf@x}{\pgf@xa-0.5*\minwidth}%
			\else
				\pgf@y=0\pgf@y
				\pgf@xa= \pgf@x
				\pgfmathsetlength{\pgf@x}{\pgf@xa-1*\minwidth}%
			\fi
		\fi
		\if\direction\direcs
			\ifnum#1=0
				\pgf@y=-\pgf@y
				\pgf@xa= \pgf@x
				\pgfmathsetlength{\pgf@x}{\pgf@xa-0.5*\minwidth}%
			\else
				\pgf@y=0\pgf@y
				\pgf@xa= \pgf@x
				\pgfmathsetlength{\pgf@x}{\pgf@xa-0*\minwidth}%
			\fi
		\fi
	\fi

	\ifnum#2=2	%
		\northeast \pgf@xa=\pgf@x \pgf@ya=\pgf@y
		\southwest \pgf@xb=\pgf@x \pgf@yb=\pgf@y

		\pgf@xc=\pgf@xa \advance\pgf@xc by -\pgf@xb	
		\pgfmathsetlength{\pgf@x}{\pgf@xa-(#1 + 0.5)*(\pgf@xc/\nport)}%
		\pgf@yc=\pgf@ya \advance\pgf@yc by -\pgf@yb	
		\pgfmathsetlength{\pgf@y}{\pgf@ya-(#1 + 0.5)*(\pgf@yc/\nport)}%

		\if\direction\direce
			\pgf@x=-\pgf@x
		\fi
		\if\direction\direcw
			\pgf@x=-\pgf@x
		\fi
	\fi
}

\pgfaddtoshape{pbsshape}{
	\anchorlet{c}{center}
	\anchorlet{n}{north}
	\anchorlet{e}{east}
	\anchorlet{s}{south}
	\anchorlet{w}{west}
	\anchorlet{se}{south east}
	\anchorlet{sw}{south west}
	\anchorlet{ne}{north east}
	\anchorlet{nw}{north west}
	\anchorlet{wsw}{west south west}
	\anchorlet{wnw}{west north west}
	\anchorlet{ene}{east north east}
	\anchorlet{ese}{east south east}
	\anchorlet{nnw}{north north west}
	\anchorlet{nne}{north north east}
	\anchorlet{ssw}{south south west}
	\anchorlet{sse}{south south east}
}

\pgfmathsetmacro{\WSSSINEHEIGHT}{0.06}

\tikzset{
	/tikz/pbskeys/.cd,
	size/.initial=0.5,
	color/.initial=O,
	direction/.initial=e,
	linestyle/.initial={linestyle},
	nport/.initial=1,
	fillgradient/.initial=O,
	/tikz/pbs/.code={
			\pgfqkeys{/tikz/pbskeys}{#1}%
			\tikzset{/tikz/pbskeys/drawer/.expanded=%
					{\pgfkeysvalueof{/tikz/pbskeys/direction}}%
					{\pgfkeysvalueof{/tikz/pbskeys/size}}%
					{\pgfkeysvalueof{/tikz/pbskeys/color}}%
					{\pgfkeysvalueof{/tikz/pbskeys/linestyle}}%
					{\pgfkeysvalueof{/tikz/pbskeys/fillgradient}}%
			}
		},
	/tikz/pbskeys/drawer/.code n args={5}{%
			\tikzset{
				pbsshape,
				minimum height=#2*\NODESIZE,
				minimum width=#2*\NODESIZE,
				#3,
				#4,
				draw,
				append after command={
						\pgfextra{\let\bdr=\tikzlastnode%
							\node[#5, fit=(\bdr.nw)(\bdr.se)] (boxgradient){};

							\if#1e
								\draw[#3, ---] ($(\bdr.nw)!.01!(\bdr.se)$) to ($(\bdr.se)!.01!(\bdr.nw)$);
							\fi
							\if#1w
								\draw[#3, ---] ($(\bdr.nw)!.01!(\bdr.se)$) to ($(\bdr.se)!.01!(\bdr.nw)$);
							\fi
							\if#1n
								\draw[#3, ---] ($(\bdr.ne)!.01!(\bdr.sw)$) to ($(\bdr.sw)!.01!(\bdr.ne)$);
							\fi
							\if#1s
								\draw[#3, ---] ($(\bdr.ne)!.01!(\bdr.sw)$) to ($(\bdr.sw)!.01!(\bdr.ne)$);
							\fi

						}
					}
			}
		},
}

\ifdefined\fontchoice
\else
	\def\fontchoice{times} 
\fi

\usepackage{pdftexcmds}

\ifnum\pdf@strcmp{\fontchoice}{firasans}=0 %
	\usepackage[book]{FiraSans} 
	\usepackage[T1]{fontenc}
	
	\tikzset{every picture/.style={/utils/exec={\sffamily}}}
\fi

\ifnum\pdf@strcmp{\fontchoice}{times}=0 %
	\usepackage{times}
\fi

\ifnum\pdf@strcmp{\fontchoice}{timesnewroman}=0 %
	\usepackage{mathptmx}
	\usepackage[T1]{fontenc}
\fi

\ifnum\pdf@strcmp{\fontchoice}{helvetica}=0 %
	\usepackage[scaled]{helvet}
	
	\usepackage[T1]{fontenc}
\fi

\makeatother
\pgfplotscreateplotcyclelist{foo}{
        {C1,mark=*},
        {C2,mark=square*},
        {C3, mark = triangle*},
        {C4,mark=+},
        {C5, mark = diamond*},
        {C6, mark = x}
      }
\begin{document}
\selectlanguage{english}    


\title{Early Termination of Low-Density Parity-Check Codes for Continuous-Variable Quantum Key Distribution \vspace{-2mm}}%



\author{Kadir G\" um\" u\c s, Jo\~ ao dos Reis Fraz\~ ao, Vincent van Vliet, Menno van den Hout, \\ Aaron Albores-Mejia, Thomas Bradley, and Chigo Okonkwo \vspace{-2mm}}

\maketitle                  

\begin{strip}
    \begin{author_descr}

        \textsuperscript{(1)} High-Capacity Optical Transmission Laboratory, Electro-Optical Communications Group, Eindhoven
University of Technology, Eindhoven, The Netherlands, 
        \textcolor{blue}{\uline{k.gumus@tue.nl}} 

            \vspace{-3mm}

    \end{author_descr}
\end{strip}

\renewcommand\footnotemark{}
\renewcommand\footnoterule{}

\begin{strip}
    \begin{ecoc_abstract}
    We analyse the impact of log a-posteriori early termination on the decoding throughput of reconciliation for continuous-variable quantum key distribution. Increases in decoded secret key rate throughput of up to 182\% are reported in both simulations and experiments. 
\textcopyright~2025 The Author(s) \vspace{-3.5mm}
    \end{ecoc_abstract}
\end{strip}

\section{Introduction}%
Continuous-variable quantum key distribution (CV-QKD) \cite{grosshans2003quantum} has garnered attention in recent years for the implementation of quantum secure key sharing using standard telecommunication components \cite{Laudenbach_2018}. CV-QKD technology has significantly matured, and commercial products have become available \cite{PETRUS}. However, the main bottleneck regarding the real-time implementation of CV-QKD systems is error correction, which limits the achievable key rates \cite{yang2023information}. 

Due to the quantum channel's low signal-to-noise ratio (SNR), complex error correction codes are required during reconciliation to ensure that the shared keys are identical between Alice and Bob. Although many experimental works on CV-QKD with high data rates exist \cite{hajomer2024experimentalcomposablekeydistribution,Zhang_2020, hajomer2025chip}, the error correction is almost always done offline. Large clusters of A100 GPUs have been used for real-time error correction \cite{wang2025high}, though this is expensive and impractical. Thus, the key rates reported are not achievable in practical systems. A popular class of error correction codes for CV-QKD are low-density parity check (LDPC) codes \cite{Milicevic_2018, gumucs2021low}. These codes are decoded using belief propagation (BP), an iterative algorithm, where the number of iterations required for the decoding directly correlates to the speed of the decoding. Hence, lowering the number of decoding iterations allows for higher secret key rate (SKR) throughputs. High-speed implementations for CV-QKD do exist, however, these works either implement min-sum decoding \cite{zhou2023high}, which is not suitable for CV-QKD because of the significant loss of performance \cite{cil2024iteration}, or focus on high-rate codes \cite{chen2024efficient}.

In this paper, we investigate using early termination (ET) based on log a-posteriori probabilities for the high frame error rate (FER) regime for low-rate codes. We show that a simple ET al algorithm can decrease the number of decoding iterations, increasing the SKR throughput by up to 182\%. We validate these results using experimental data and show that decoded key rates almost triple.

\vspace{-3mm}
\section{Early Termination}
The BP algorithm used for LDPC decoding is an iterative algorithm that involves message passing along a bipartite graph, which is defined by the parity check matrix $\mathbf{H}$. Every iteration, the belief of each bit is updated until the decoder has reached the maximum number of iterations $D_{max}$. For the low-rate LDPC codes used for CV-QKD, $D_{max}$ is in the range of 200-500 \cite{hajomer2024experimentalcomposablekeydistribution,Milicevic_2018}. One way to reduce the number of iterations is to use ET. 

A popular method of ET is based on the calculation of the syndrome of the codeword estimate $\mathbf{\hat{c}}$, which has to be equal to 0, i.e., $\mathbf{\hat{c}}\mathbf{H} = \mathbf{0}$ \cite{Milicevic_2018}. In this case, the decoder has converged to a valid codeword, and thus, the decoder output will not change. This method works well when the FER is low. However, for the high FER regime, using the parity check equation (PCE) ET will often allow the decoding to continue until $D_{max}$ has been reached, as the decoder does not converge.

In the high FER regime, we can use an ET method based on the log a-posteriori probabilities of the bits after each iteration \cite{Kienle2005}, which is referred to as variable node reliability early termination (VNR-ET). Let $l_{out,i}^ {(k)}$ be the log a-posteriori probability of bit $i$ after $k$ iterations. For each iteration, we calculate $q^{(k)} = \sum\limits_{i\in \mathcal{V}}^{} l_{out,i}^ {(k)}$, where $\mathcal{V}$ denotes the set of variable nodes with degree higher than 1. Most variable nodes in low-rate LDPC codes are of degree 1, and these do not have to be considered when calculating $q^{(k)}$. For most correctly decoded codewords $q^{(k)}$ increases monotonically, i.e., $q^{(k)} > q^{(k-1)}$. With VNR-ET, we stop decoding if at any point $q^{(k)} < q^{(k-1)}$. Compared to the calculations of the $tanh$ and $atanh$ functions, this step adds a negligible amount of complexity \cite{Kienle2005}.

\begin{figure*}[t!]
\begin{subfigure}{0.49\linewidth}
      \begin{tikzpicture}[>=latex]
\begin{axis}[
every axis/.append style={font=\small},
tick label style={font=\footnotesize},
xlabel = $\beta$(\%),
ylabel = FER,
xmin = 92, xmax =100,
ymin = 0.01, ymax = 1,
ymode = log,
xtick distance=1,
x tick label style={yshift= -1mm},
y tick label style={xshift= -1mm},
ylabel shift = 1mm,
xlabel shift = 1mm,
set layers, mark layer=axis tick labels,
grid = major,
width = \linewidth,
 height = 6cm,
 xlabel near ticks,  
 ylabel near ticks,  
 xticklabel style={/pgf/number format/fixed},
every axis plot/.append style={thick},legend style={at={(0.5,0.23)},anchor=west, font = \scriptsize,row sep=-0.75ex,inner sep=0.2ex,fill opacity = 0.6, text opacity = 1},
legend cell align={left},
cycle list name = foo
]

\pgfplotstableread{Figures/FER.txt}
\datatable
\pgfplotsinvokeforeach {1,3,4,5,6}{
\addplot+
         table
         [
          x expr=\thisrowno{0}, 
          y expr=\thisrowno{#1} 
         ] {\datatable};
}
\addlegendentry{$D_{max} = 1000$}
\addlegendentry{$D_{max} = 500$}
\addlegendentry{$D_{max} = 250$}
\addlegendentry{$D_{max} = 100$}
\addlegendentry{VNR-ET}

\end{axis}
\end{tikzpicture}
\end{subfigure}
\begin{subfigure}{0.49\linewidth}
      \begin{tikzpicture}[>=latex]
\begin{axis}[
every axis/.append style={font=\small},
tick label style={font=\footnotesize},
xlabel = $\beta$(\%),
ylabel = $\overline{D}$,
xmin = 92, xmax =100,
ymin = 0, ymax = 1000,
xtick distance=1,
x tick label style={yshift= -1mm},
y tick label style={xshift= -1mm},
ylabel shift = 1mm,
xlabel shift = 1mm,
set layers, mark layer=axis tick labels,
grid = major,
 xlabel near ticks,  
 width = \linewidth,
  height = 6cm,
 ylabel near ticks,  
 xticklabel style={/pgf/number format/fixed},
every axis plot/.append style={thick},legend style={at={(0.1,0.75)},anchor=west, font = \scriptsize,row sep=-0.75ex,inner sep=0.2ex,fill opacity = 0.6, text opacity = 1},
legend cell align={left},
cycle list name = foo
]

\pgfplotstableread{Figures/Iterations.txt}
\datatable
\pgfplotsinvokeforeach {1,3,4,5,6}{
\addplot+
         table
         [
          x expr=\thisrowno{0}, 
          y expr=\thisrowno{#1} 
         ] {\datatable};
}
\addlegendentry{$D_{max} = 1000$}
\addlegendentry{$D_{max} = 500$}
\addlegendentry{$D_{max} = 250$}
\addlegendentry{$D_{max} = 100$}
\addlegendentry{VNR-ET}

\end{axis}
\end{tikzpicture}
\end{subfigure}
\caption{Simulation: For different amounts of $D_{max}$ and VNR-ET. Left: $\beta$ vs. FER, Right: $\beta$ vs. $\overline{D}$.}
\label{fig:FER}
\end{figure*}
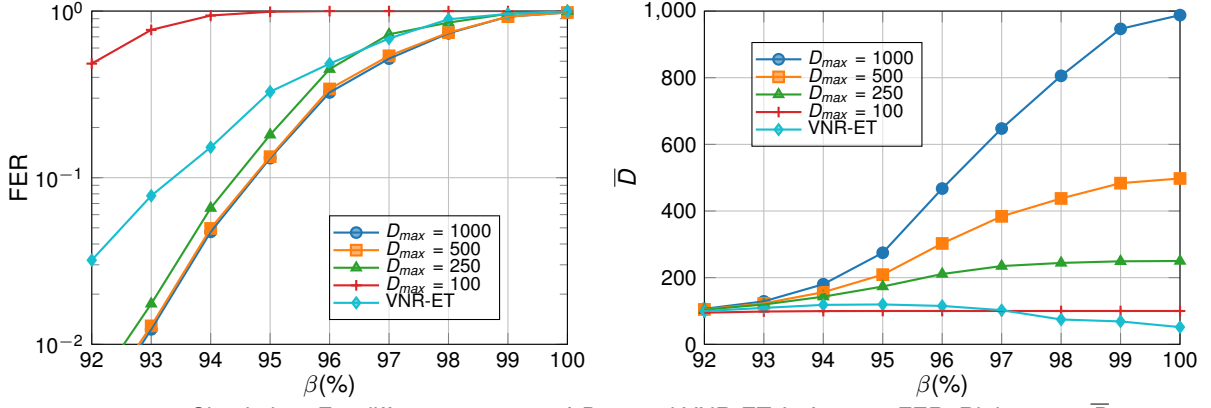

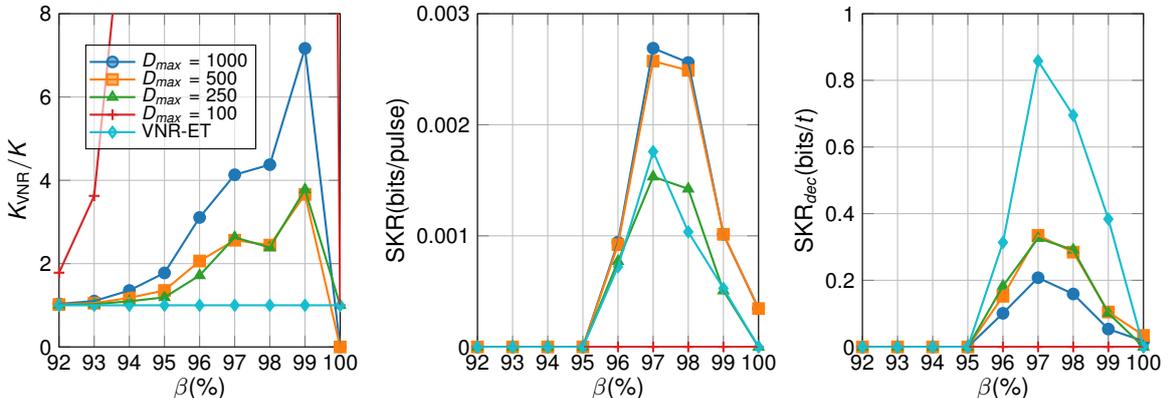
\begin{figure*}[!b]
\centering
    \begin{subfigure}{0.33\linewidth}
      \begin{tikzpicture}[>=latex]
\begin{axis}[
every axis/.append style={font=\small},
tick label style={font=\footnotesize},
xlabel = $\beta$(\%),
ylabel = $K_{\text{VNR}}/K$,
xmin = 92, xmax =100,
ymin = 0, ymax = 8,
xtick distance=1,
x tick label style={yshift= -1mm},
y tick label style={xshift= -1mm},
ylabel shift = 1mm,
xlabel shift = 1mm,
set layers, mark layer=axis tick labels,
grid = major,
width = \linewidth,
 height = 6cm,
 xlabel near ticks,  
 ylabel near ticks,  
 xticklabel style={/pgf/number format/fixed},
every axis plot/.append style={thick},legend style={at={(0.7,0.75)},anchor=east, font = \scriptsize,row sep=-0.75ex,inner sep=0.2ex,fill opacity = 0.6, text opacity = 1},
legend cell align={left},
cycle list name = foo
]

\pgfplotstableread{Figures/Relative_Throughput_Gain.txt}
\datatable
\pgfplotsinvokeforeach {1,3,4,5,6}{
\addplot+
         table
         [
          x expr=\thisrowno{0}, 
          y expr=\thisrowno{#1} 
         ] {\datatable};
}
\addlegendentry{$D_{max} = 1000$}
\addlegendentry{$D_{max} = 500$}
\addlegendentry{$D_{max} = 250$}
\addlegendentry{$D_{max} = 100$}
\addlegendentry{VNR-ET}
\end{axis}
\end{tikzpicture}
    \end{subfigure}
    \hspace{-5mm}
    \begin{subfigure}{0.33\linewidth}
          \begin{tikzpicture}[>=latex]
\begin{axis}[
every axis/.append style={font=\small},
tick label style={font=\footnotesize},
xlabel = $\beta$(\%),
ylabel = SKR(bits/pulse),
xmin = 92, xmax =100,
ymin = 0, ymax = 3,
xtick distance=1,
 yticklabels={0,$0$,$0.001$,$0.002$,$0.003$},
x tick label style={yshift= -1mm},
y tick label style={xshift= -1mm},
ylabel shift = 1mm,
xlabel shift = 1mm,
set layers, mark layer=axis tick labels,
grid = major,
width = \linewidth,
 height = 6cm,
 xlabel near ticks,  
 ylabel near ticks,  
 xticklabel style={/pgf/number format/fixed},
every axis plot/.append style={thick},legend style={at={(0.55,0.8)},anchor=east, font = \scriptsize,row sep=-0.75ex,inner sep=0.2ex,fill opacity = 0.6, text opacity = 1},
legend cell align={left},
cycle list name = foo
]

\pgfplotstableread{Figures/SKR_finitesize.txt}
\datatable
\pgfplotsinvokeforeach {1,2,3,4,5}{
\addplot+
         table
         [
          x expr=\thisrowno{0}, 
          y expr=\thisrowno{#1} 
         ] {\datatable};
}
\end{axis}
\end{tikzpicture}
    \end{subfigure}
    \begin{subfigure}{0.33\linewidth}
          \begin{tikzpicture}[>=latex]
\begin{axis}[
every axis/.append style={font=\small},
tick label style={font=\footnotesize},
xlabel = $\beta$(\%),
ylabel = SKR$_{dec}$(bits/$t$),
xmin = 92, xmax =100,
ymin = 0, ymax = 1,
xtick distance=1,
x tick label style={yshift= -1mm},
y tick label style={xshift= -1mm},
ylabel shift = 1mm,
xlabel shift = 1mm,
set layers, mark layer=axis tick labels,
grid = major,
 xlabel near ticks,  
 width = \linewidth,
 height = 6cm,
 ylabel near ticks,  
 xticklabel style={/pgf/number format/fixed},
every axis plot/.append style={thick},legend style={at={(0.05,0.8)},anchor=west, font = \scriptsize,row sep=-0.75ex,inner sep=0.2ex},
legend cell align={left},
cycle list name = foo
]

\pgfplotstableread{Figures/Throughput.txt}
\datatable
\pgfplotsinvokeforeach {1,2,3,4,5}{
\addplot+
         table
         [
          x expr=\thisrowno{0}, 
          y expr=\thisrowno{#1} 
         ] {\datatable};
}

\end{axis}
\end{tikzpicture}
    \end{subfigure}
    \caption{Simulation: For different $D_{max}$, and for VNR-ET. Right:  $\beta$ vs. $K_{VNR-ET}/K$, Middle: SKR vs. $\beta$, Left: $\text{SKR}_{dec}$ vs. $\beta$}
    \label{fig:SKR_Sim}
\end{figure*}

\vspace{-1.6mm}
\section{Secret Key Rate}
The SKR of a CV-QKD system is often expressed in bits/pulse and is given by \cite{Milicevic_2018}:
\begin{equation}
    \label{SKR}
    \text{SKR} = (1-\text{FER})(\beta I_{AB} - \chi_{BE} - \Delta_n),
\end{equation}
where $I_{AB}$ is the mutual information between Alice and Bob, $\chi_{BE}$ is the Holevo information, and $\Delta_n$ is the finite-size penalty of the privacy amplification block size and depends on $N_{privacy}$\cite{Milicevic_2018}. The reconciliation efficiency $\beta$ indicates how close the error correction is to the capacity and is defined as $R/I_{AB}$, where $R$ is the rate of the code. There is a trade-off between $\beta$ and FER, so the highest $\beta$ will not always result in the highest SKR \cite{eriksson2020digital}. The SKR throughput in bits/s is $f_{rep}\text{SKR}$, where $f_{rep}$ is the pulse rate of the system. 

The key rate related to the practical implementation of a CV-QKD system depends on the throughput of the error correction, which is smaller than SKR throughput in state-of-the-art works \cite{yang2023information,hajomer2025chip}. The throughput $K$ of an LDPC decoder in bits/$t$, where $t$ is the latency per iteration, can be expressed as: 
\begin{equation}
    K = \frac{N}{\overline{D}}R(1-\text{FER}), 
    \label{Throughput}
\end{equation}
where $N$ is the blocklength of the code and $\overline{D}$ is the average number of decoding iterations. The latency $t$ is hardware-dependent and is not relevant for this analysis.

By substituting Eq.~\eqref{Throughput} into Eq.~\eqref{SKR}, the decoded SKR throughput in bits/$t$ can be written as:
\begin{equation}
     \text{SKR}_{dec} = \frac{N}{\overline{D}\mu}(1-\text{FER})(\beta I_{AB} - \chi_{BE} - \Delta_n).
\end{equation}
where $\mu = 1$ if homodyne detection is used, and $\mu = 2$ if heterodyne detection is used. For the rest of the paper, heterodyne detection is assumed.

\begin{figure*}[t!]
    \centering
    \resizebox{\linewidth}{!}{\input{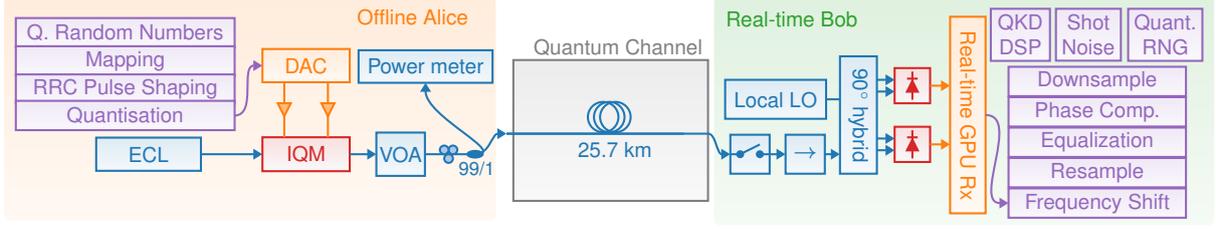}}
    \caption{Experimental set-up of CV-QKD over a 25.7 km fibre, with offline Alice and real-time GPU-based Bob}
    \label{fig:experiment}
    \vspace{-1mm}
\end{figure*}

\begin{figure*}[b!]
\vspace{-3mm}

      \begin{tikzpicture}[>=latex]
\begin{axis}[
every axis/.append style={font=\small},
tick label style={font=\footnotesize},
xlabel = Capture Number,
ylabel = SKR$_{dec}$(bits/$t$),
xmin = 1, xmax =33,
ymin = 0, ymax = 40,
xtick distance=1,
axis y line* = left,
xticklabel style = {xshift=0.23cm},
xticklabels = {0,1,2,3,...,32},
x tick label style={yshift= -1mm},
y tick label style={xshift= -1mm},
ylabel shift = 1mm,
xlabel shift = 1mm,
set layers, mark layer=axis tick labels,
ymode = log,
width = \linewidth,
 height = 5cm,
 xlabel near ticks,  
 ylabel near ticks,  
 xticklabel style={/pgf/number format/fixed},
every axis plot/.append style={thick},legend style={at={(0.1,0.5)},anchor=west, font = \scriptsize,row sep=-0.5ex,inner sep=0.2ex,fill opacity = 0.6, text opacity = 1},
legend cell align={left},
cycle list name = foo
]

\fill[C8,opacity = 0] (axis cs: 1,0.001) rectangle (axis cs: 2,100);
\fill[C8,opacity = 0.1] (axis cs: 2,0.001) rectangle (axis cs: 12,100);
\fill[C8,opacity = 0.2] (axis cs: 12,0.001) rectangle (axis cs: 13,100);
\fill[C8,opacity = 0.3] (axis cs: 13,0.001) rectangle (axis cs: 26,100);
\fill[C8,opacity = 0.4] (axis cs: 26,0.001) rectangle (axis cs: 29,100);
\fill[C8,opacity = 0.5] (axis cs: 29,0.001) rectangle (axis cs: 33,100);

\pgfplotstableread{Figures/SKR_Experiment.txt}
\datatable
\pgfplotsinvokeforeach {1,3}{
\addplot+[only marks]
         table
         [
          x expr=\thisrowno{0}, 
          y expr=\thisrowno{#1} 
         ] {\datatable};
}
\addlegendentry{$\text{SKR}_{dec,\text{VNR-ET}}$}
\addlegendentry{$\text{SKR}_{dec,D_{max} = 250}$}
\addlegendimage{only marks, C4, mark = x}
\addlegendentry{$\text{SKR}_{dec,\text{VNR-ET}}/\text{SKR}_{dec,D_{max} = 250}$}
\end{axis}

\begin{axis}[
every axis/.append style={font=\small},
tick label style={font=\footnotesize},
ylabel = $\text{SKR}_{dec,\text{VNR}}/\text{SKR}_{dec,D_{max} = 250}$,
xmin = 1, xmax =33,
ymin = 1, ymax = 3,
xtick = \empty,
x tick label style={yshift= 1mm},
y tick label style={xshift= 1mm},
ylabel shift = 1mm,
xlabel shift = 1mm,
set layers, mark layer=axis tick labels,
axis y line* = right,
width = \linewidth,
 height = 5cm,
 xlabel near ticks,  
 ylabel near ticks,  
 xticklabel style={/pgf/number format/fixed},
every axis plot/.append style={thick},legend style={at={(0.13,0.5)},anchor=west, font = \scriptsize,row sep=-0.75ex,inner sep=0.2ex},
legend cell align={left},
cycle list name = foo
]

\pgfplotstableread{Figures/SKR_Experiment.txt}
\datatable
\addplot[only marks, C4, mark = x]
         table
         [
          x expr=\thisrowno{0}, 
          y expr=\thisrowno{4} 
         ] {\datatable};
\end{axis}

\node at (0.23,3.6){\footnotesize 92};
\node at (2.7,3.6){\footnotesize 93};
\node at (5.17,3.6){\footnotesize 94};
\node at (8.33,3.6){\footnotesize 95};
\node at (11.92,3.6){\footnotesize 96};
\node at (13.5,3.6){\footnotesize 97};
\node at (7.3,4){$\beta_{opt}(\%)$};
\end{tikzpicture}
    \caption{Experiment: The SKR$_{dec}$ for different CV-QKD captures comparing $D_{max} = 250$ to VNR-ET. For error correction, an expanded TBP-LDPC code with $R = \frac{1}{5}$ punctured to the appropriate rate is used. The captures are sorted according to at which $\beta$ SKR$_{dec}$ was maximised for VNR-ET. Darker backgrounds correspond to higher $\beta_{opt}$.}
    \label{fig:SKR_Experiment}
\end{figure*}
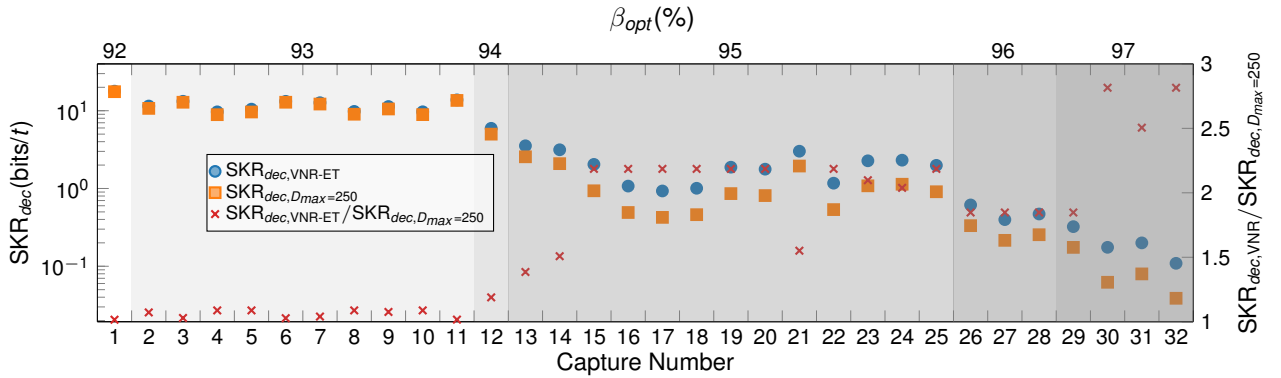 

\vspace{-1.6mm}
\section{Simulations}
 For the simulations in this section, we use a $R= \frac{1}{50}$ type-based protograph (TBP) LDPC code \cite{gumucs2021low} with blocklength $N = 10^6$. We have simulated for four different amounts of maximum decoding iterations $D_{max} = \{100, 250, 500, 1000\}$ and for when using VNR-ET. in the case of VNR-ET, the $D_{max}$ is unbounded, although during the simulations $D$ never exceeded 300. In all cases, we use PCE-ET. For the CV-QKD system we assume the following variables: Gaussian modulation with modulation variance $V_A$ chosen such that $I_{AB} = \frac{R}{\beta}$, heterodyne detection, quantum efficiency $\eta = 0.4$, electronic noise $\nu_{el} = 0.01$, excess noise on Bob $\xi_{Bob} = 0.001$, transmission distance $d = 80$ km over an optical fibre with loss $\alpha = 0.2$ dB/km, and privacy amplification blocklength $N_{privacy} = 10^8$.
 
In Fig. \ref{fig:FER}, we show FER (left) and $\overline{D}$ (right) against the reconciliation efficiency $\beta$. The FER curve is the same for when $D_{max} = 500$ and when $D_{max} = 1000$, indicating that more than 500 decoding iterations is unnecessary for this particular code. For $D_{max} = 250$ the decrease in FER compared to $D_{max} = 500$ is manageable; however, when we set $D_{max}$ to 100, the performance decrease is very significant. In general, we found that when $D_{max} < 250$, the reduction in FER performance is too substantial to justify the decrease in $D_{max}$. For VNR-ET, we also get a penalty in FER, as we sometimes stop decoding a codeword which would have eventually converged. The average number of iterations when using VNR-ET is much lower when $\beta$ is high, as the FER is high as well. When $\beta$ decreases, the FER decreases; hence, the number of codewords at which VNR-ET would have an effect decreases as well. As a result, for lower $\beta$, $\overline{D}$ is similar when using VNR-ET or not. 

In Fig. \ref{fig:SKR_Sim} (left), we compare the ratio of $K$ to $K_{VNR-ET}$, which is the throughput in the case of VNR-ET. As is shown, $K_{VNR-ET}$ is always larger, despite the higher FER, except at $\beta = 100\%$. Here, the FER for VNR-ET is  1, while for $D_{max} \geq 500$, the FER is very close to 1 instead. Otherwise, as $\beta$ increases, the throughput for VNR-ET increases. 

In Fig. \ref{fig:SKR_Sim} (middle) and (right), we compare both SKR and SKR$_{dec}$, respectively, for the CV-QKD system previously described. The SKR for VNR-ET is smaller than for $D_{max} \geq 500$, but similar to $D_{max} = 250$. However, for SKR$_{dec}$, the throughput for VNR-ET outperforms the other curves, with a gain in SKR$_{dec}$ of 182\%. This gain depends on the optimal reconciliation efficiency $\beta_{opt}$ of the CV-QKD system, as is shown in the next section. 

\vspace{-2mm}
\section{Experimental Results}
Figure \ref{fig:experiment} shows our practical CV-QKD implementation. The offline transmitter (Alice) uses an external cavity laser (ECL) with a linewidth $<100$~kHz, IQ optical modulation for probabilistically shaped 256-QAM signals, and a 250~Mbaud symbol rate with 50\% pilot symbols. Alice's average $V_A$ was between 7 to 8 SNUs.

On Bob's side, a receiver operates in calibration and quantum signal reception mode by using an optical switch. Key system capabilities include an ECL as a local oscillator (LO) into an optical hybrid, digitisation at 2~GS/s, and real-time digital signal processing (DSP) for calibration and quantum signal analysis. DSP includes frequency-offset compensation, filtering, equalisation, pilot-based phase recovery \cite{Sjoerd2023}, and parameter estimation. 

Bob follows the trusted noise assumption, and all losses due to coupling and the optical isolator are considered in the quantum efficiency, resulting in a decrease in $\eta$ from 67\% to 40\%. The average $\xi_{Bob}$ is 0.0056 SNU, and we have a clearance of 10~dB. An isolator between the optical switch and the hybrid prevents back reflections from the local LO. Real-time post-processing on a GPU evaluates security via excess noise and SKR analyses. 

For the error correction, we use an $R = \frac{1}{5}$ expanded TBP-LDPC code punctured to the appropriate rate with $N = 10^5$. For each capture $\beta$ was optimised to maximise SKR. Figure \ref{fig:SKR_Experiment} shows the SKR for 32 different captures, comparing VNR-ET to only PCE-ET with $D_{max} = 250$. The captures have been sorted according to $\beta_{opt}$ for VNR-ET. This fluctuates over the captures as the excess noise of the system fluctuates as well. The gains are small ( < 10\%) for cases where $\beta_{opt} \leq 93\%$. However, when $\beta_{opt} = 97\%$, SKR$_{dec}$ is almost tripled. SKR$_{dec}$ decreases as $\beta_{opt}$ increases, as these are the captures where $I_{AB}$ and $\chi_{BE}$ are closer to each other. VNR-ET would therefore offer a bigger improvement for systems which have to operate at high $\beta$, which would correspond more to CV-QKD systems with lower-grade components. 

\vspace{-2mm}
\section{Conclusion}
In this work, we have shown how applying VNR-ET during decoding increases SKR throughputs because of the higher throughput of the decoder. We have demonstrated increases in SKR$_{dec}$ of up to 182\% in both simulations and experiments.
\newpage
\clearpage
\section{Acknowledgements}
This work was supported by the PhotonDelta GrowthFunds Programme on Photonics and by the Dutch Ministry of Economic Affairs (EZ), as part of the Quantum Delta NL KAT-2 programme on Quantum Communications. This work also acknowledges the support of the European Union via the Marie Curie Doctoral Network QuNEST (Grant Agreement: 101120422) and the EIC Transition Grant project PAQAAL (under Grant Agreement 101213884)

\printbibliography

\end{document}